\documentclass[prb,showpacs,twocolumn,amsmath,amssymb]{revtex4}

\usepackage{graphicx}
\usepackage{psfrag}
\usepackage{dcolumn}
\usepackage{bm}

\begin{document}

\newcommand{\ba}{\begin{array}}
\newcommand{\ea}{\end{array}}
\newcommand{\bc}{\begin{center}}
\newcommand{\ec}{\end{center}}
\newcommand{\nen}{\nonumber}
\newcommand{\eps}{\epsilon}
\newcommand{\dbar}{{\delta \!\!\!\! \smallsetminus}}
\newcommand{\tr}{\textrm{ tr}}
\newcommand{\sumint}{{\textstyle \sum}\hspace{-2.7ex}\int}
\newcommand{\atanh}{\textrm{atanh}}

\newcommand{\UA}{\uparrow}
\newcommand{\DA}{\downarrow}

\newcommand{\x}{\mathbf{x}}
\newcommand{\ist}{\!=\!}
\newcommand{\ket}[1]{{\left| #1 \right>}}
\newcommand{\bra}[1]{{\left< #1 \right|}}
\newcommand{\bracket}[2]{{\left< {#1} \left| {#2} \right.\right>}}
\newcommand{\bigbracket}[2]{{\big< {#1} \big| {#2} \big>}}
\newcommand{\Bigbracket}[2]{{\Big< {#1} \Big| {#2} \Big>}}
\newcommand{\bigket}[1]{{\big| #1 \big>}}
\newcommand{\bigbra}[1]{{\big< #1 \big|}}
\newcommand{\Bigket}[1]{{\Big| #1 \Big>}}
\newcommand{\Bigbra}[1]{{\Big< #1 \Big|}}
\newcommand{\me}[3]{{\left< {#1} \left| {#2} \right| {#3} \right>}}
\newcommand{\bigme}[3]{{\big< {#1} \big| {#2} \big| {#3} \big>}}
\newcommand{\bigmeS}[3]{{\big< {#1} \big| {#2} \big| {#3} \big>_S}}
\newcommand{\Bigme}[3]{{\Big< {#1} \Big| {#2} \Big| {#3} \Big>}}
\newcommand{\green}[1]{{ \left<\!\left< {#1} \right>\!\right> }}
\newcommand{\biggreen}[1]{{ \big<\!\big< {#1} \big>\!\big> }}
\newcommand{\Biggreen}[1]{{ \big<\!\big< {#1} \big>\!\big> }}

\newcommand{\define}{\stackrel{\textrm{\tiny def}}{=}}
\newcommand{\mustbe}{\stackrel{!}{=}}
\newcommand{\n}{\Hat{n}}
\newcommand{\cl}{{\cal L}}
\newcommand{\phd}{^{\phantom{\dagger}}}
\newcommand{\SC}{{\cal S}}
\newcommand{\Dint}{\int {\cal D}[\SC]\;}
\newcommand{\YY}{{\cal Z}}

\newcommand{\cdag}{c^{\dagger}}
\newcommand{\cnod}{c^{\phantom{\dagger}}}
\newcommand{\fdag}{f^{\dagger}}
\newcommand{\fnod}{f^{\phantom{\dagger}}}
\newcommand{\bdag}{b^{\dagger}}
\newcommand{\bnod}{b^{\phantom{\dagger}}}
\newcommand{\Ueff}{{U_{\textrm{eff}}}}
\newcommand{\tUeff}{{\tilde U_{\textrm{eff}}}}
\renewcommand{\bottomfraction}{0.99}
\renewcommand{\textfraction}{0.01}


\title{Dynamic response functions for the Holstein-Hubbard model}

\author{W. Koller}\email{w.koller@imperial.ac.uk}
\author{D. Meyer}\email{d.meyer@imperial.ac.uk}
\author{A. C. Hewson}\email{a.hewson@imperial.ac.uk}
\affiliation{Department of Mathematics, Imperial College, London SW7 2BZ, UK
}

\date{30 March 2004}

\begin{abstract}
We present results on the dynamical correlation functions of the
particle-hole symmetric Holstein-Hubbard model at zero temperature,
calculated using the dynamical mean field theory which is solved by the
numerical renormalization group method. 
We clarify the competing influences of the electron-electron and
electron-phonon interactions particularity at the 
different metal to insulator transitions.
The Coulomb repulsion is found to dominate the behaviour in large parts of the
metallic regime. By suppressing charge fluctuations, it effectively decouples
electrons from phonons.
The phonon propagator shows a characteristic softening near the metal to
bipolaronic transition but there is very little softening on the approach to the
Mott transition.
\end{abstract}



\maketitle

\section{Introduction}

Phonons are important in metallic systems.
They affect the electronic behaviour in diverse ways causing the dominant
temperature dependent contribution to the resistivity, an enhancement
of the specific heat, and an attractive effective electron-electron
interaction, which may induce superconductivity.
One can expect strong electron-phonon interactions in strongly correlated
systems, such as heavy fermion compounds\cite{FKZ88}, where the radius of the
rare earth ions is very sensitive to the $f$-electron occupation.
The coupling of electronic states to the lattice also plays an important role
in the anomalous electronic behaviour of the
manganites\cite{MLS95,IFT98,Edw02}.
The superconductivity in the fullerides\cite{Gun97} appears to be induced by
electron-phonon interactions within a strongly correlated electron band.

However, the fully quantum mechanical treatment of lattice effects in these
compounds has so far received comparatively little attention.
The reason for that lies in the difficulty in handling strong
electron-phonon coupling together with strong electron-electron repulsion.
Recent advances in non-perturbative techniques, most notably the
development of the dynamical mean field theory (DMFT)\cite{GKKR96},
enable one to investigate the interplay of electron and phonon
interactions in these systems.

One of the most prominent models in the field of strongly
correlated electron systems is the Hubbard model\cite{Hub63} which
been used extensively to study the effects of strong local
electron-electron interactions.
Especially within the framework of DMFT\cite{GKKR96} the nature of the
Mott transition could be clarified\cite{RZK92,Jar92}.
The Holstein model\cite{Hol59} has been used to study polaronic effects in the
absence of electron-electron repulsion, mainly in the limit of low
electron density.
Systems with finite electron density have received less
attention, but have been studied within the DMFT using Monte
Carlo\cite{FJS93}, exact diagonalization (ED)\cite{CC03}, perturbative
approaches\cite{Fre94,FJ94,HD01pre} and the numerical renormalization group
(NRG)\cite{MHB02}.

Even less studied has been the interplay of electron-electron interactions and
the electron-phonon coupling\cite{FJ95,HG00,HG00b,CSCCG04,DM02a}.
The most natural model incorporating both these terms is the
Holstein-Hubbard model defined by the Hamiltonian
\begin{equation}                                        \label{eq:hamil}
  \begin{aligned}
    H =&
    \sum_{\vec{k}\sigma}
    \epsilon(\vec{k})\, \cdag_{\vec{k}\sigma} \cnod_{\vec{k}\sigma} +
    U \sum_{i} n_{i\uparrow} n_{i\downarrow} \\ &+
    \omega_0 \sum_i  \bdag_i \bnod_i +
    g\sum_i  (\bdag_i + \bnod_i) \big(n_{i\UA}+ n_{i\DA}-1 \big) \:.
    \end{aligned}
\end{equation}
Here $U$ describes the electron--electron interaction within a
band of dispersion $\epsilon(\vec{k})$.
The electron density $n_{i\UA} + n_{i\DA}$ at site $i$ couples linearly to
the local displacement operator
$x_i \equiv (\bnod_i + \bdag_i)/\sqrt{2  m\omega_0}$ 
with an electron-phonon coupling $g$. 
The phonons are assumed to be dispersionless (local Einstein phonons) with
energy $\omega_0$ and $m$ is the mass of the vibrating ions.

\begin{figure}
  \includegraphics[width=0.46\textwidth]{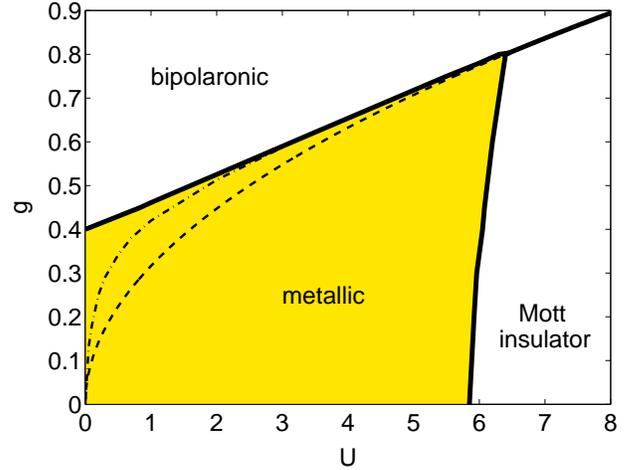}
  \caption{Zero temperature phase diagram of the half filled Holstein-Hubbard
  model for   $\omega_0=0.2$ and a semielliptic band of width $W=4$ as
  calculated in   Ref.~\onlinecite{KMOH03pre}.
  The dashed line is the locus of $U_{\textrm{eff}}=0$.
  Points on the dot-dashed line fulfil $\big<n_\UA n_\DA\big>=1/4$.}
  \label{phases.eps}
\end{figure}

Recently, the $T=0$ phase diagram of the half filled
Holstein-Hubbard model has been calculated\cite{KMOH03pre,JPHLC03pre} using
various local approximations, and is shown in Fig.~\ref{phases.eps}.
The results are for a semi-elliptic band of width $W=4$ and phonon
frequency $\omega_0=0.2$ ($m=1$).
All types of long-range order are excluded.
We can distinguish three different phases: metallic, bipolaronic
and Mott insulating phase.
The metallic phase is always found to be a Fermi liquid.
The latter two phases display a gap in the one-electron spectra and will be
referred to as gapped phases.
The parameters of Fig.~\ref{phases.eps}, namely $W=4, m=1$ and $\omega=0.2$
are taken throughout this paper, unless mentioned otherwise.

One way of simplifying the discussion of the Holstein-Hubbard model
would be to integrate out the phonons in order to derive an effective
model for the electrons.
The effective action is then governed by the $\omega$-dependent
potential\cite{HM02}
\begin{equation}                                        \label{eq:Ueff_def}
  \Ueff(\omega) = U + \frac{2\, g^2\, \omega_0}{\omega^2-\omega_0^2}\:.
\end{equation}
This reduces to the static quantity
$\Ueff \equiv U -2\,g^2/\omega_0$ in the limits $\omega \to 0$ and
$\omega_0 \to \infty$.
Therefore, on small energy scales, we expect an effective attractive
or repulsive interaction depending on the value of $\Ueff$.
For small phonon energies $\omega_0$, however, there is no reason to expect
the static quantity $\Ueff$ to be sufficient to characterize the behaviour of
the Holstein-Hubbard model.
The properties of this model are expected to depend on both $U$ and $g$
individually.
This is also apparent from the phase diagram in Fig.~\ref{phases.eps}.
The dot-dashed line, where the double occupancy takes the value of a
free system $\big<n_\UA n_\DA\big>=1/4$ is quite distinct from the line
$\Ueff=0$.
The aim of this study is to clarify the interplay of these different types of
interactions.

An interesting question about this model is whether the suppression
of local charge fluctuations by a significant Hubbard $U$ will effectively
decouple electrons and phonons on all energy scales.
It is possible, however, that for the behaviour on low energy
scales the effective onsite-$U$ is strongly renormalized and hence the
electron phonon coupling will have a comparatively stronger effect on energy
scales $\omega < \omega_0$  (see Ref~\onlinecite{HG00b, FJ95}).

In this paper, we present results on dynamical electronic and phonon
response functions for the Holstein-Hubbard model as calculated within the
dynamical mean field theory\cite{GKKR96} (DMFT).
The effective impurity problem is solved using Wilson's numerical
renormalization group (NRG) method\cite{KWW80a} extended to treat the
electron--phonon coupling\cite{HM02}.
The combination of DMFT and NRG has proven to be a reliable and effective
method for calculations of both metallic and insulating phases at $T=0$.
It has been applied to the Mott transition in the pure Hubbard
model\cite{Bul99} and to the pure Holstein model\cite{MHB02}.
In our calculations we use the discretization parameter $\Lambda=1.8$
and retain up to $1200$ states.
The value of $\Lambda=1.8$ gives rise to a bandwidth correction factor
$A_\Lambda=1.029$ (see Eq. (5.42) in \onlinecite{KWW80a}) which we
take into account in the NRG procedure.
For the calculation of dynamical response functions we use the method
described in Ref.~\onlinecite{BHP98}.

Apart from the usual local one-electron Green's function
$G_\sigma(\omega)\equiv\biggreen{\cnod_{i\sigma};\cdag_{i\sigma}}_\omega$
we will be interested in two different phonon Green's functions,
\begin{equation}                                             \label{eq:dD_def}
  \begin{aligned}
    d(\omega) &= \biggreen{\bnod_i\,;\bdag_i}_\omega\:,\\
    D(\omega) &= 2\omega_0\,\biggreen{x_i\,;x_i}_\omega =
    \biggreen{\bnod_i+\bdag_i\,;\bnod_i+\bdag_i}_\omega\:.
  \end{aligned}
\end{equation}
Since we will treat only  homogeneous phases here and exclude all
types of long-range order, we have dropped the site indices in all of
the definitions.
The spectral weight of $d(\omega)$ will be denoted by
$\rho_d(\omega) \equiv - \frac{1}{\pi}\, \textrm{Im}
\,d(\omega+i0^+)$,  and similar for all other Green's functions.
The dynamical charge and spin susceptibility are given by
\begin{equation}                                           \label{eq:chi_csss}
  \begin{aligned}
    \chi_c(\omega) &= \biggreen{\hat O \,; \hat O}_\omega\:,\\
    \chi_s(\omega) &= \biggreen{n_{i\UA}-n_{i\DA}; n_{i\UA}-n_{i\DA}}_\omega\:,
  \end{aligned}
\end{equation}
respectively.
The operator $\hat O$ represents the electronic term that couples to the
phonons,
\begin{equation}                                           \label{eq:O_def}
  \hat O = n_{i\UA} + n_{i\DA} - 1\:.
\end{equation}
It should be noted that our definition of $\chi_s(\omega)$ differs from the
usual one by a factor of four.
We also calculate the quasiparticle weight $z$ given by
$z = (1-\textrm{Re}\,\Sigma'(0))^{-1}$ within the local approximation
of the DMFT.

The different regimes of the phase diagram of Fig.~\ref{phases.eps}
give rise to the following structure of the paper:
In Sec.~\ref{sec:g_dependence} we choose a fixed value of $U$ and study
dynamical properties as a function of the phonon coupling $g$.
We find continuous transitions from a metallic to a bipolaronic state
for small values of $U=0$ and $U=1$ (Sec.~\ref{ssec:weakU}).
For a larger value of $U=5$, as discussed in Sec.~\ref{ssec:strongU},
the transition to the bipolaronic state becomes discontinuous.
Complementarily, we look at the $U$-dependence of dynamical properties in
Sec.~\ref{sec:U_dependence} keeping $g$ fixed to $g=0.45$.
For this value of $g$ the system can be in all three phases, depending
on $U$.
In Sec~\ref{sec:polaronic} we study the properties of the model on the
two polaronic lines as given by $\Ueff=0$ and
$\langle n_\UA n_\DA\rangle = 1/4$ (dashed and dot-dashed lines in
Fig.~\ref{phases.eps}).
We also show spectra for $\Ueff=0$ for different values of $\omega_0$.
In Sec.~\ref{sec:overview} we give an overview of our results, and we
present our conclusions in Sec.~\ref{sec:conclusions}.


\section{Dependence on $g$ -- phonon driven phase transition}
                                                        \label{sec:g_dependence}

In this section, we fix the value of the Hubbard interaction~$U$ and study
dynamical properties for a varying electron-phonon coupling $g$. We
distinguish between the cases of  weak and  strong $U$.

\subsection{Weak $U$}                                      \label{ssec:weakU}

%
\begin{figure}
  \includegraphics[width=0.46\textwidth]{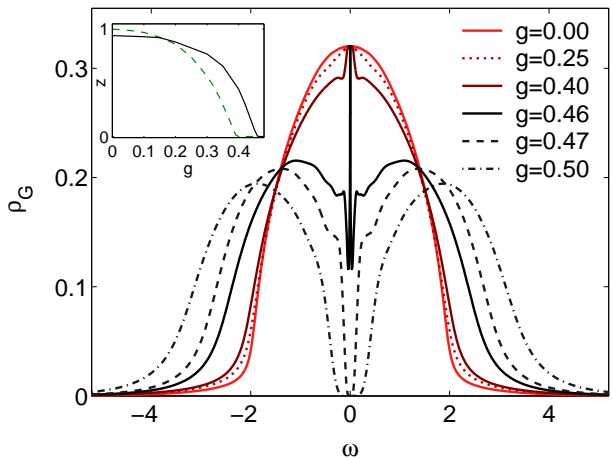}
  \caption{One-particle spectral function for $U=1$ and various
  values of $g$. The inset shows the variation of the quasiparticle
  weight $z$ for $U=1$ (solid line) and $U=0$ (dashed line).}
  \label{g_spec_U10_gx.eps}
\end{figure}

First of all, we look at the one-electron spectra for a fixed $U = 1$ as shown
in Fig.~\ref{g_spec_U10_gx.eps}. There is a strong similarity of these results
with those of the pure Holstein model ($U=0$) calculated
previously\cite{MHB02}. At weak coupling a narrow peak emerges at the Fermi
energy. The top of this feature is rather flat and the imaginary part of the
self energy has a shallow $\omega^2$-dependence.  This one would expect from
lowest order perturbation theory in $g$, where the imaginary part of the self
energy vanishes in the range $|\omega|<\omega_0$ for $U=0$ (see Eq.~(19) in
Ref.~\onlinecite{HM02}). Upon increasing $g$ further, there is a rapid
narrowing of this feature until it disappears at a critical coupling of
$g_c\approx 0.47$ and a gap opens. Similar to the pure Holstein model, there
is no preformed gap in this case in contrast to the Mott transition in the
Hubbard model.

The metallic regime corresponds to a renormalized Fermi liquid. The
quasiparticle weight $z$ is shown as a function of $g$ in the inset together
with that for the pure Holstein model ($U=0$). For $U=1$, the quasiparticles
are already slightly renormalized at $g=0$. However, initially the
renormalization with $g$ is weaker than in the case $U=0$ but more rapid on
the approach to $g_c$\cite{KMOH03pre}.
As with the pure Holstein model, there is no evidence of multiple solutions
near the phase transition. This is in contrast to the situation near the Mott
transition for the Hubbard model, where there is a finite coexistence region.

\begin{figure}
  \includegraphics[width=0.23\textwidth]{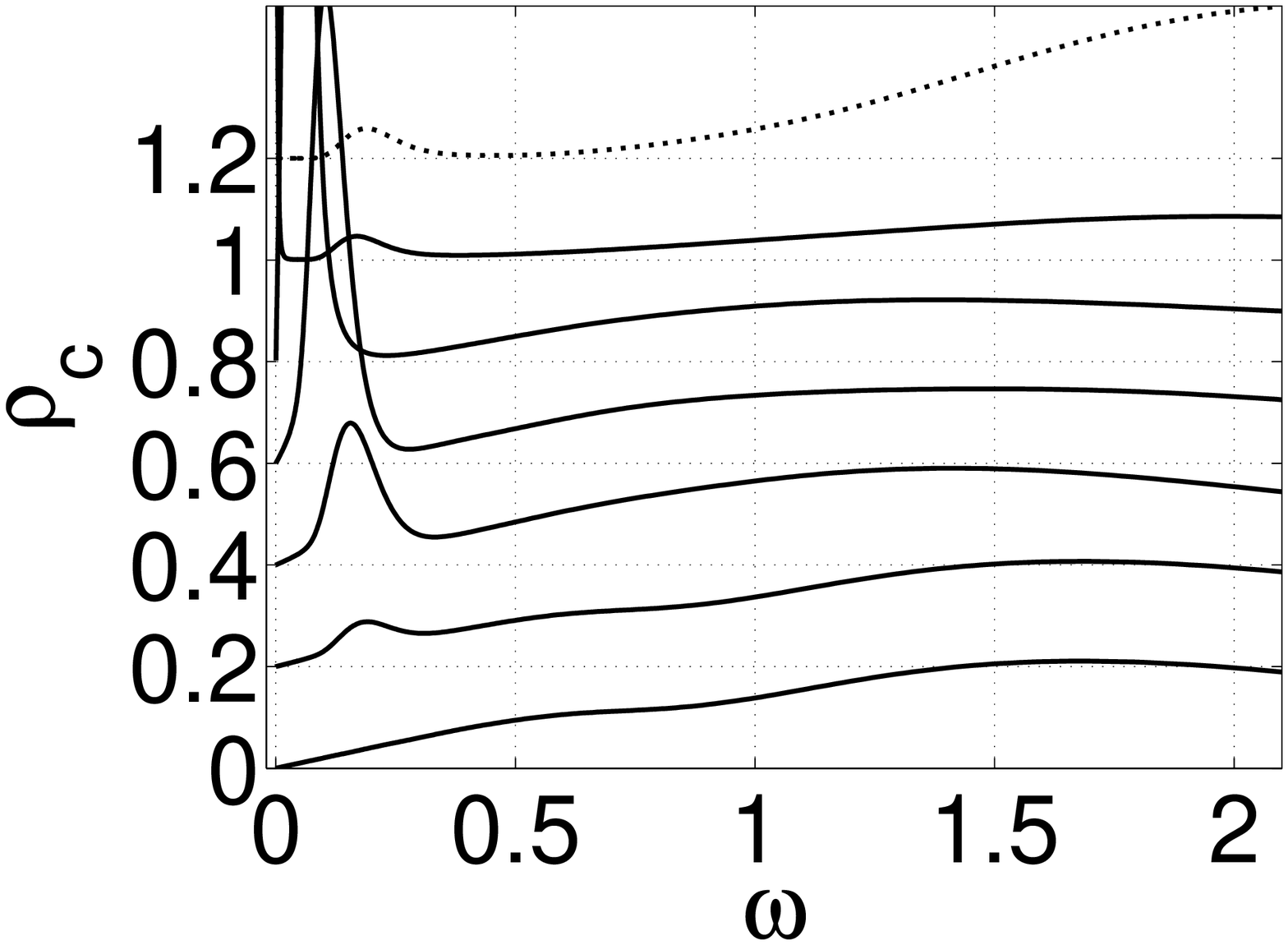}
  \includegraphics[width=0.23\textwidth]{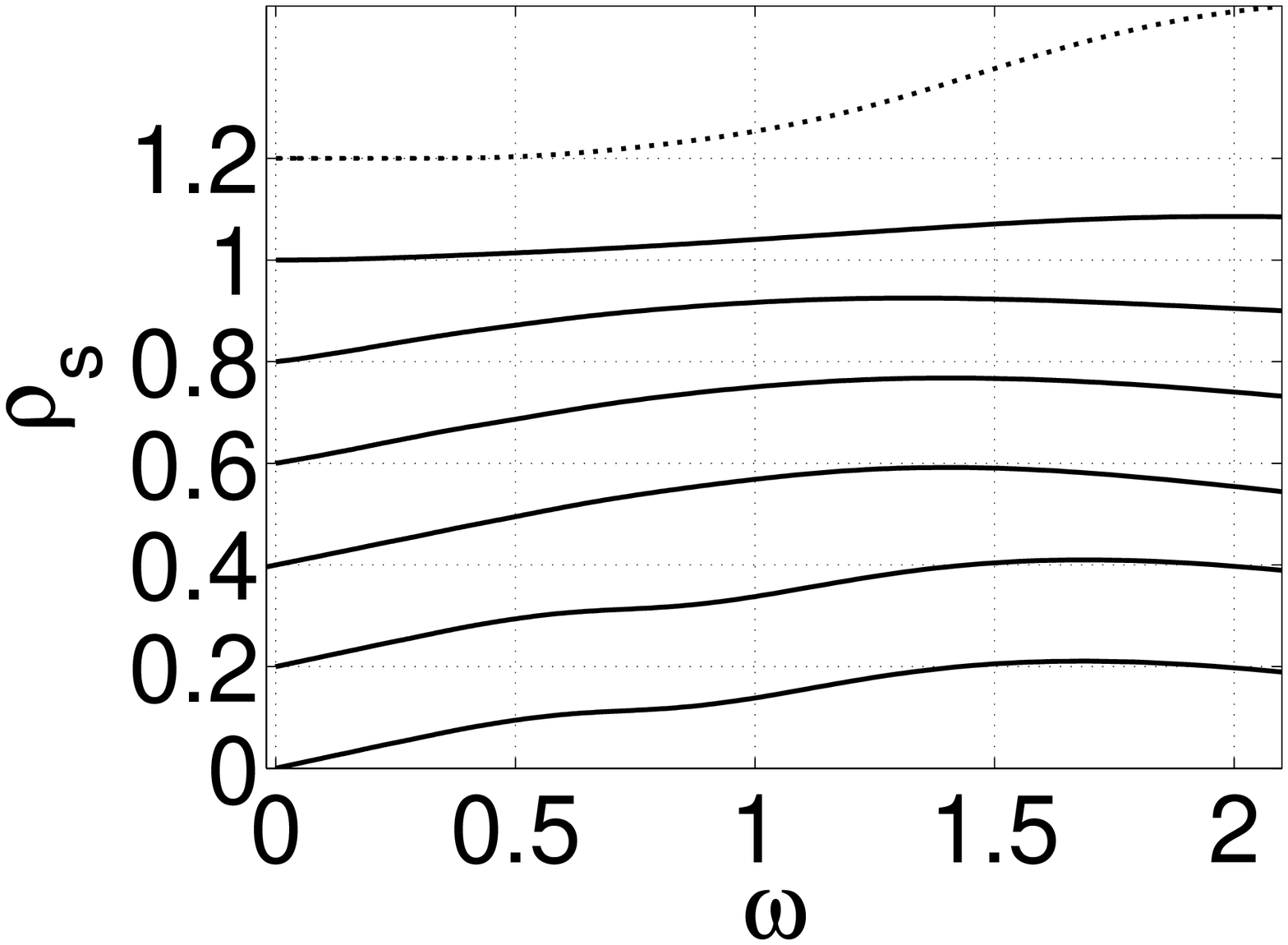}\\
  \includegraphics[width=0.23\textwidth]{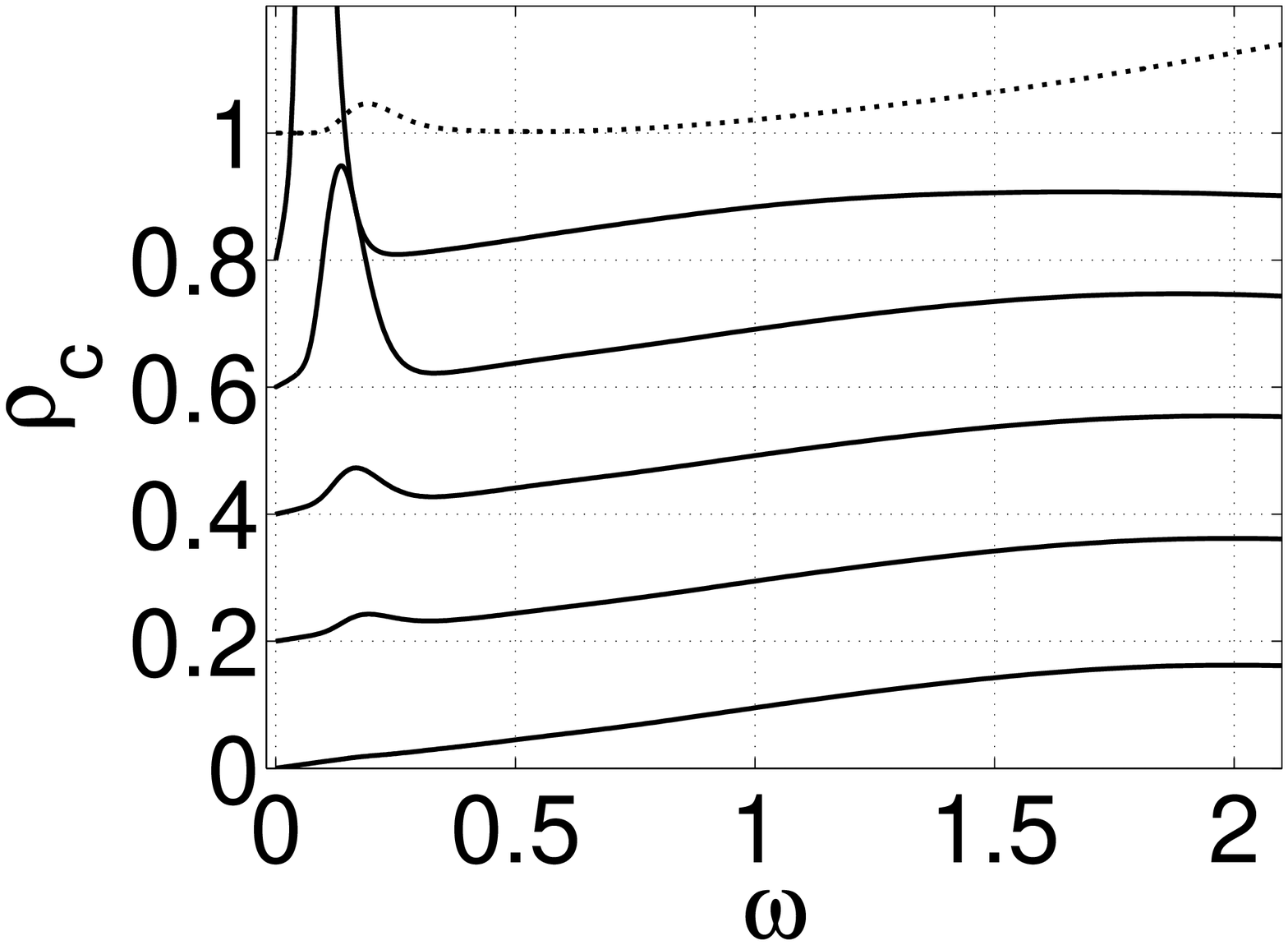}
  \includegraphics[width=0.23\textwidth]{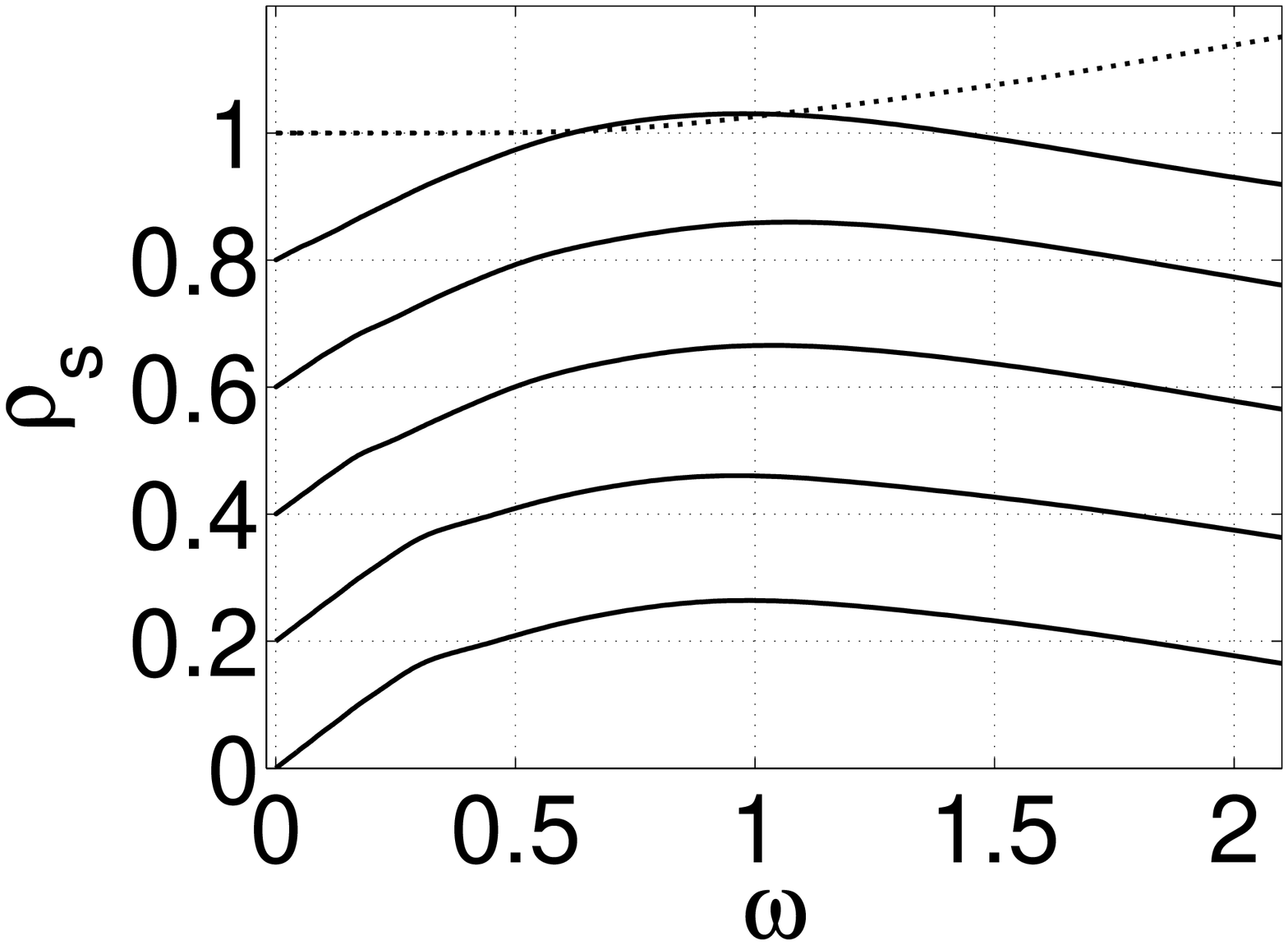}
  \caption{Low energy behaviour of the spectra of the charge (left)
    and spin-susceptibility (right) for the pure Holstein model
    ($U=0$, top) and $U=1$ (bottom).
    The values of $g$ are $g=0.0$ (bottom), $g=0.1, 0.2, 0.3, 0.37, 0.40,
    0.42$ (top) in the upper panel and $g=0.0$ to $0.5$ in steps of $0.1$ in
    the lower panel. The dotted lines are scaled up by a factor of $10$.}
  \label{csss_spec_U10_gx.eps}
\end{figure}

Figure~\ref{csss_spec_U10_gx.eps} shows the spectra of the corresponding
charge- and spin susceptibilities at $U=1$ and $U=0$ for comparison.
In the free system, $\chi_c(\omega)=\chi_s(\omega)$ as defined in
Eq.~(\ref{eq:chi_csss}), which can be seen in the lowest curves of the two
upper panels. 
Looking at the case $U=0$ first (upper panel), we see the emergence and
buildup  of a low-energy peak in $\chi_c$ and the suppression of weight in
$\chi_s$ for $\omega<W/2$.
In the bipolaronic state ($g=0.42$), both charge and spin susceptibilities
have the same gap, roughly twice that seen in the electronic spectrum.
They have an identical peak above this gap. These can be seen in the dotted
curves of Fig.~\ref{csss_spec_U10_gx.eps}, which have been scaled up by a
factor of $10$ to make these features visible.
The coincidence of these peaks can be explained by the fact that spin
fluctuations necessarily require the breakup of a bipolaronic pair.
Their weight is very small due to the strong bipolaronic binding.
There is an additional very small peak in the charge susceptibility in the gap
close to $\omega_0$ from the residual couplings to the phonon mode.
In the absence of charge ordering, there must be charge fluctuations at
$\omega=0$ which, however, cannot be observed numerically. 

For $U=1$ (lower panel) we observe the same overall trends. However, starting
from a reduced charge susceptibility due to the finite $U$, the peak in
$\chi_c$ develops more slowly. The spin susceptibility, which is enhanced due
to the finite $U$, changes very little in the metallic phase. The low-energy
spin fluctuations are almost completely suppressed with the emergence of the
gap in the bipolaronic state.

\begin{figure*}
  \begin{center}
  \includegraphics[width=0.4\textwidth]{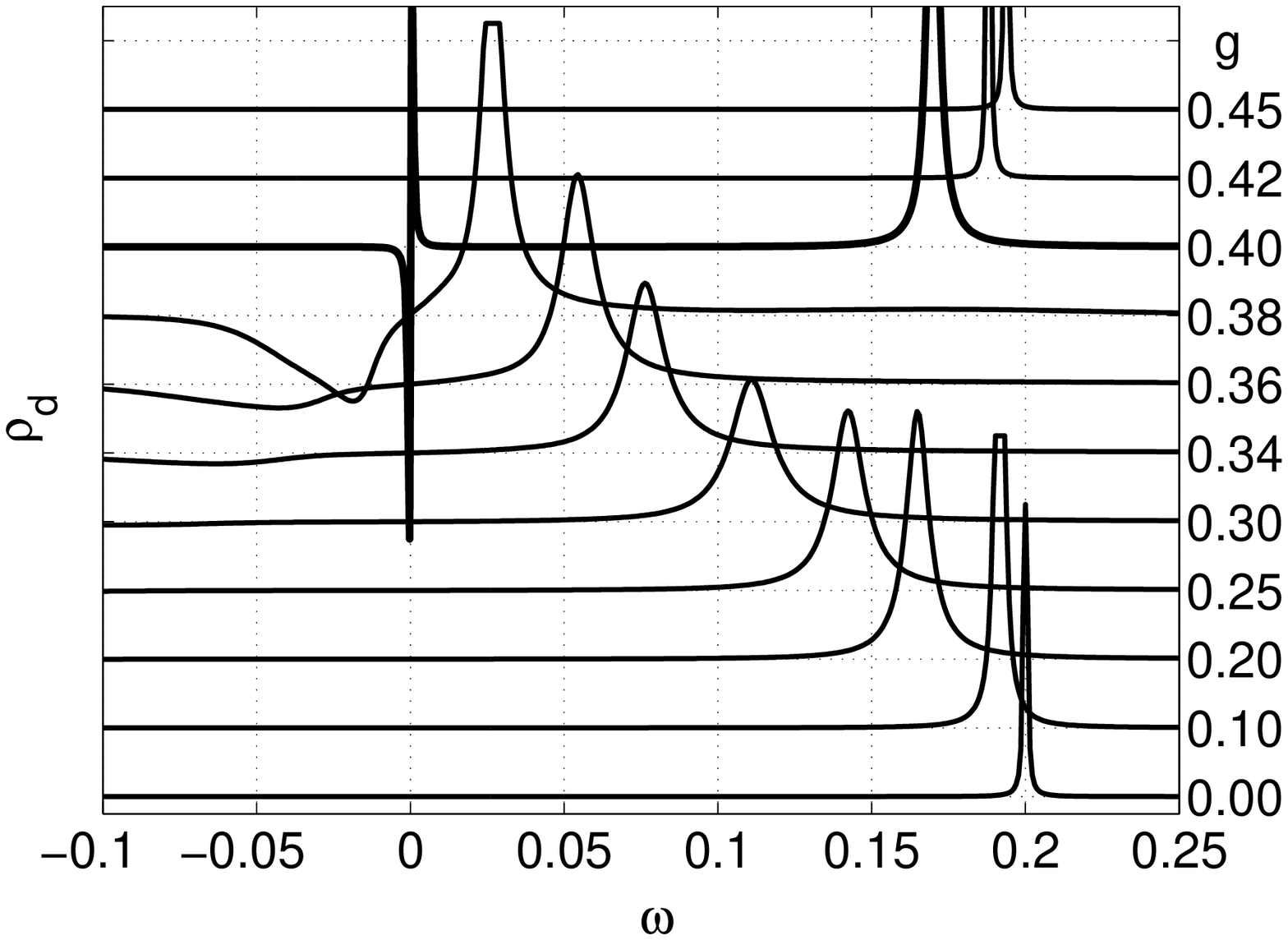}
  \hspace{2em}
  \includegraphics[width=0.4\textwidth]{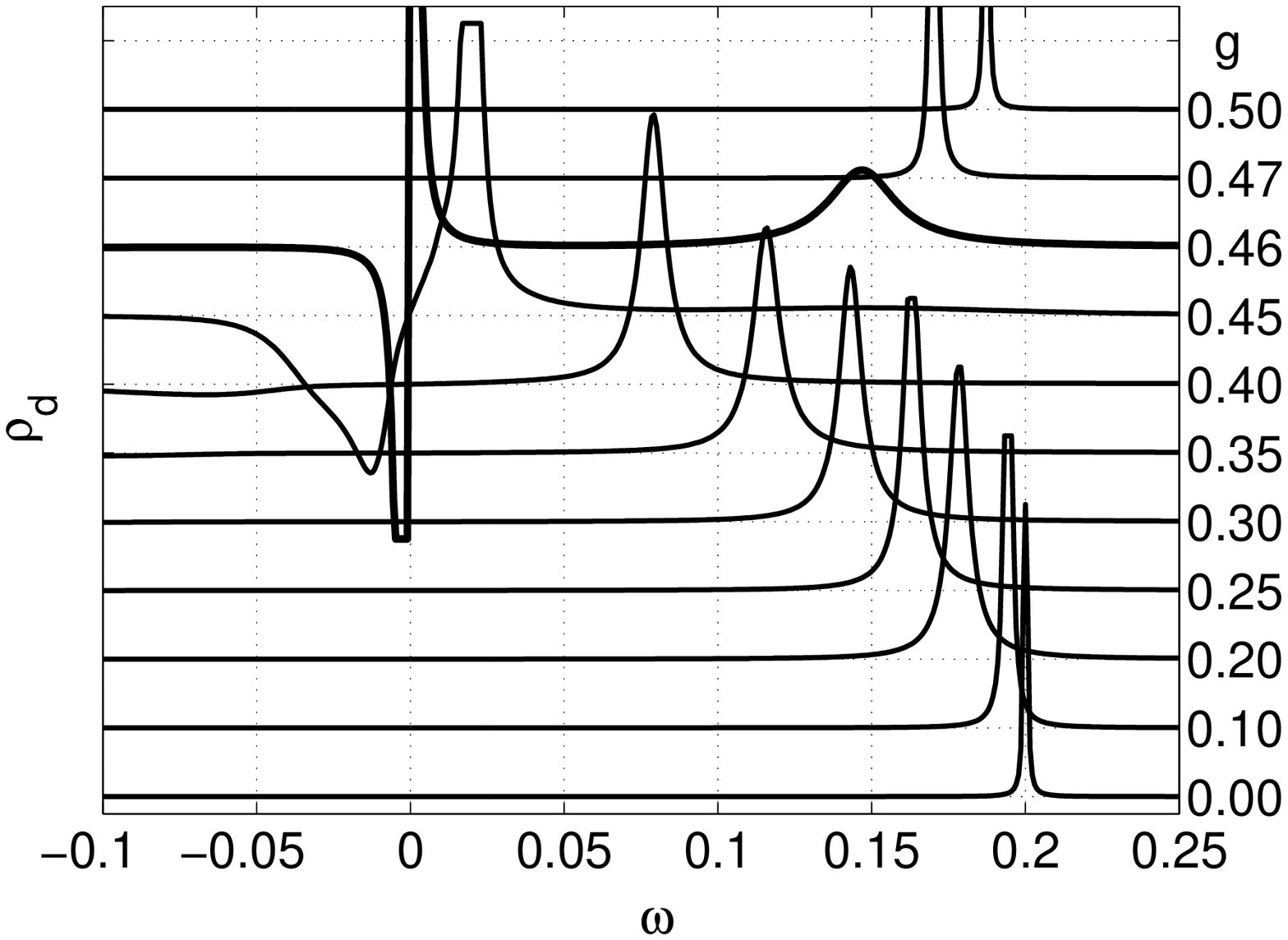}
  \end{center}
  \caption{Spectral density of the phonon propagator for $U=0$ (left) and
  $U=1$ (right) and various values of $g$.}
  \label{bb_spec_U10_gx.eps}
\end{figure*}

The charge fluctuations can be directly related to the phonon propagator via
the equation (see appendix \ref{app:d_chic})
\begin{equation}                                           \label{eq:d_chic}
  d(\omega) = d_0(\omega) + g^2\,d_0(\omega) \, \chi_c(\omega) \, d_0(\omega)\:,
\end{equation}
where $d_0(\omega) = 1/(\omega-\omega_0)$ is the free phonon propagator.
However, use of this equation with an {\em approximate} form of
$\chi_c(\omega)$ can lead to severe numerical errors, as discussed in
appendix~\ref{app:d_chic}.
In Fig.~\ref{bb_spec_U10_gx.eps}, we give the results for the phonon spectra
for $U=0$ and $U=1$ calculated via the self energy\cite{JPC03}, similar
to the procedure we use for the electronic spectra\cite{BHP98}. The results
for $U=0$ differ significantly from those published earlier\cite{MHB02} based
on Eq.~\ref{eq:d_chic}.

In contrast to the earlier results, the main phonon peak is found initially to
soften significantly in a similar way to the Migdal Elisahberg result.
The two-peak structure develops only upon approaching the transition to the
bipolaronic state. In the gapped state, the high-frequency peak narrows and
tends towards $\omega_0$.
The same trend can be seen for $U=1$ (right-hand plot). The slower initital
softening correlates with the suppression of charge fluctuations seen in
$\chi_c(\omega)$.

\begin{figure}[b]
 \includegraphics[width=0.46\textwidth]{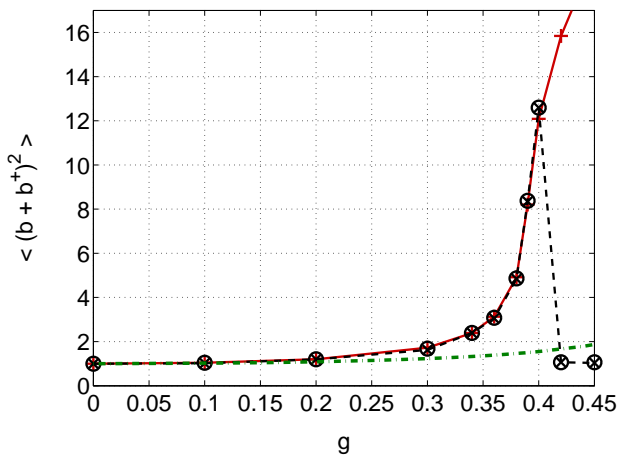}
  \caption{Expectation value of the lattice fluctuations as a function
  of the electron-phonon coupling $g$ in the pure Holstein model ($U=0$).
  In the metallic range ($g<0.4$), the  direct calculation ($+$) agrees well
  with the integration over the phonon Green's function ($\circ$) and charge
  susceptibility ($\times$)).
  For details see appendices~\ref{app:d_chic} and~\ref{app:n_vs_x}.
  The dot-dashed line shows the result from the Migdal Eliashberg calculation.}
  \label{x2_av_U00_gx.eps}
\end{figure}

On the approach to the transition, significant negative spectral weight for
$\omega<0$ can be seen in the spectra shown in
Fig.~\ref{bb_spec_U10_gx.eps}. As a consequence of the spectral theorem, this
relates directly to the average number of excited phonons in the ground
state. 
The spectra of the Green's function $D(\omega)$ show similar features as
$d(\omega)$, and integrated up to $\omega=0$ yield the average lattice
fluctuations as shown in Fig.~\ref{x2_av_U00_gx.eps}.
These results agree well with the direct evaluation of the expectation value
$\langle (b + b^\dagger)^2 \rangle$ in the metallic phase.
For low to intermediate values of $g$, the NRG results do not deviate
significantly from the Midgal Eliashberg calculation and are significantly
smaller than reported in Ref.~\onlinecite{MHB02} based on
Eq.~\ref{eq:d_chic}.
On the approach to the transition, there is a large increase in the
lattice fluctuation.
In the gapped phase, we observe a large difference between the direct
calculation and the integration over the negative part of the phonon spectral
density. One way of looking at it is that due to the limited numerical
resolution we miss the contribution from the peak at $\omega=0^-$ in
the spectrum of $D(\omega)$.
An other way of explaining this difference is that there is no
dynamics connecting the two degenerate ground states with
$\langle n\rangle = 0$ and $2$ in the gapped phase.
These two ground states have nonvanishing average displacements
$\pm x_0$, see appendix~\ref{app:n_vs_x}, and integration over the
phonon spectrum yields the fluctuations about $\pm x_0$, i.e.,
$\langle (x-x_0)^2 \rangle$. 
The calculation of $x_0$ in one of the ground states quantitatively explains
the different values of the two methods in the gapped phase.


\subsection{Strong $U$}                                 \label{ssec:strongU}

%
\begin{figure}
  \includegraphics[width=0.46\textwidth]{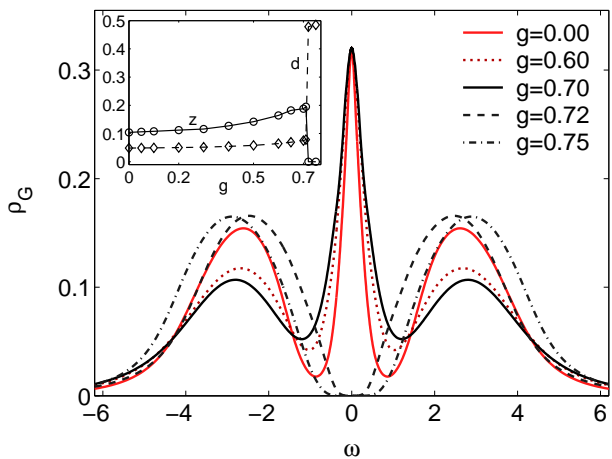}
  \caption{One-particle spectral function for $U=5$ and various
  values of $g$. The central peak broadens with increasing $g$ and at
  $g\approx 0.71$ disappears in a first order phase transition to the
  bipolaronic state. The inset shows the quasiparticle weight $z$ ($\circ$)
  and the double occupancy $d=\langle n_\UA n_\DA \rangle$ ($\diamond$) as
  functions of $g$.}
  \label{g_spec_U50_gx.eps}
\end{figure}

Next we look at the transition to the bipolaronic state for a larger fixed
value of $U=5$.
Figure~\ref{g_spec_U50_gx.eps} shows the corresponding one-particle
electron spectral functions for various values of $g$.
Their $g$-dependence is in sharp contrast to the small $U$ case shown in
Fig.~\ref{g_spec_U10_gx.eps}.
Initially for $g=0$, we have the three-peak structure of a strongly
correlated Hubbard model.
With increasing~$g$, the central resonance broadens slightly as is
reflected in the increasing quasiparticle weight $z$ (see inset). This can be
ascribed to the partial cancellation of the Hubbard repulsion by phonon
mediated electron-electron attraction, see Eq.~(\ref{eq:Ueff_def}).
The effect, however, is much weaker than one would deduce from
$\Ueff$. The position of the high-energy Hubbard bands is virtually unaffected
by the phonons, as expected from Eq.~(\ref{eq:Ueff_def}) evaluated at $\omega
\approx \pm U/2$.
When approaching the critical coupling $g_c\approx 0.72$, the
quasiparticle peak disappears abruptly and $z$ jumps to zero,
indicating a discontinuous transition to the bipolaronic phase.
The double occupancy, shown in the inset to Fig.~\ref{g_spec_U50_gx.eps}, is
much less than a quarter in the metallic state but makes a sudden jump to
$d\approx 1/2$ on the point of the phase transition. A coexistence of
metallic and gapped solutions is found in the narrow range $0.71
\lesssim g \lesssim 0.72$.

\begin{figure}[b]
  \includegraphics[width=0.23\textwidth]{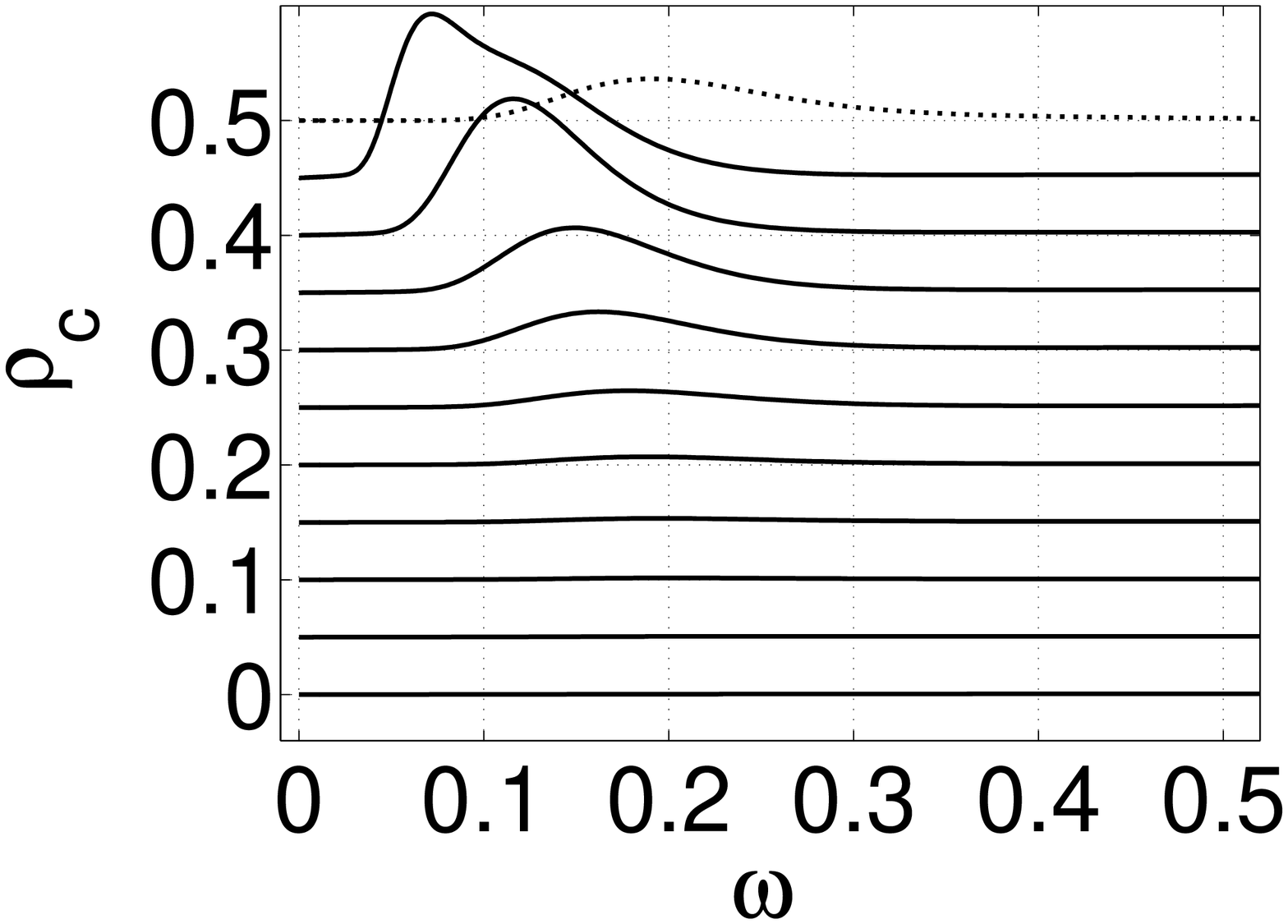}
  \includegraphics[width=0.23\textwidth]{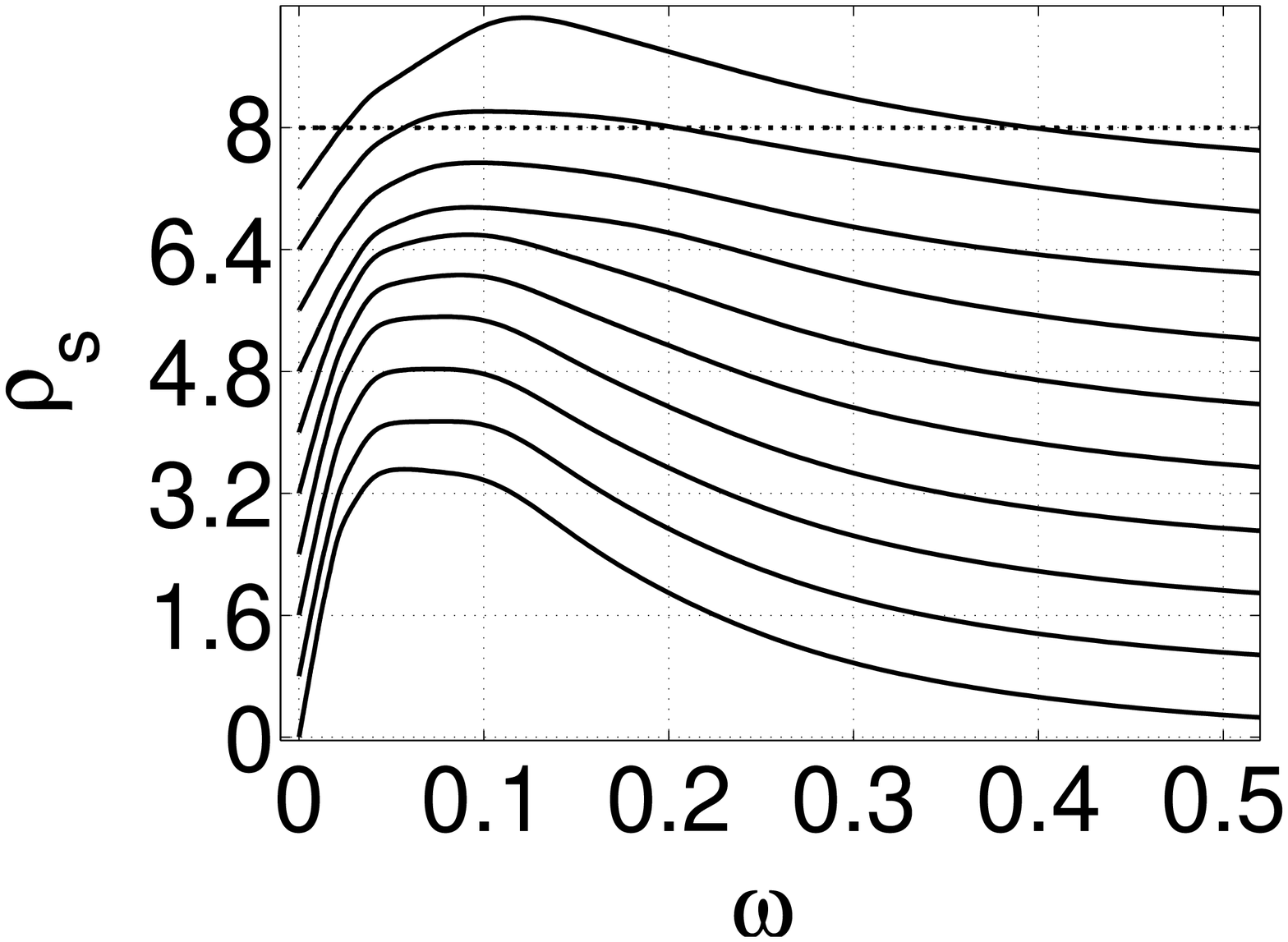}
  \caption{Low-energy behaviour of the spectra of the charge (left) and
  spin-susceptibility (right) for $U=5$.
  The values of $g$ correspond to those in  Fig.~\ref{bb_spec_U50_gx.eps}.
  Note the different scales of the $y$-axes. Dotted lines have been scaled up
  by a factor of $10$.}
  \label{csss_spec_U50_gx.eps}
\end{figure}

The spectra of the charge and spin susceptibilities are shown in
Fig.~\ref{csss_spec_U50_gx.eps}. They display similar trends, but more
pronounced, than the corresponding results for $U=1$ (lower panels of
Fig.~\ref{csss_spec_U10_gx.eps}).
The low-energy peak in $\chi_c(\omega)$ only becomes clearly visible when $g$
reaches the value of $g \approx 0.65$.
In the bipolaronic phase, the peaks above the gap in the charge and spin
susceptibilities are not identical as in the $U=0$ case, but very similar.
However, the gaps are too large to be visible on the $\omega$-scale of
Fig.~\ref{csss_spec_U50_gx.eps}. 
The only peak visible, when enhanced by a factor of $20$, is the peak in the
charge susceptibility within the gap near $\omega_0$.

\begin{figure}
  \includegraphics[width=0.46\textwidth]{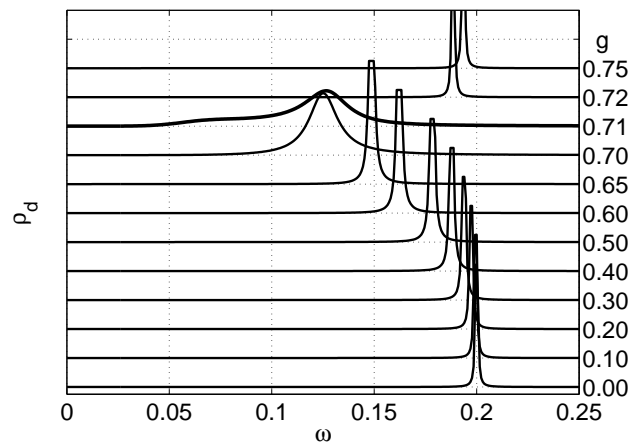}
  \caption{Spectral density of the phonon propagator for $U=5$ and various
  values of $g$.}
  \label{bb_spec_U50_gx.eps}
\end{figure}

In Fig.~\ref{bb_spec_U50_gx.eps}, the spectra of the corresponding phonon
propagators are plotted. Over a large initial range of $g\lesssim 0.6$
there is very little softening, corresponding to a complete
suppression of charge fluctuations. 
When approaching $g_c$, the phonon mode softens and develops a low-energy
shoulder. After the discontinuous transition to the bipolaronic phase, the
mode hardens back to $\omega_0$ and narrows.


\section{Dependence on $U$ -- Mott transition}          \label{sec:U_dependence}

\begin{figure*}
  \includegraphics[width=0.46\textwidth]{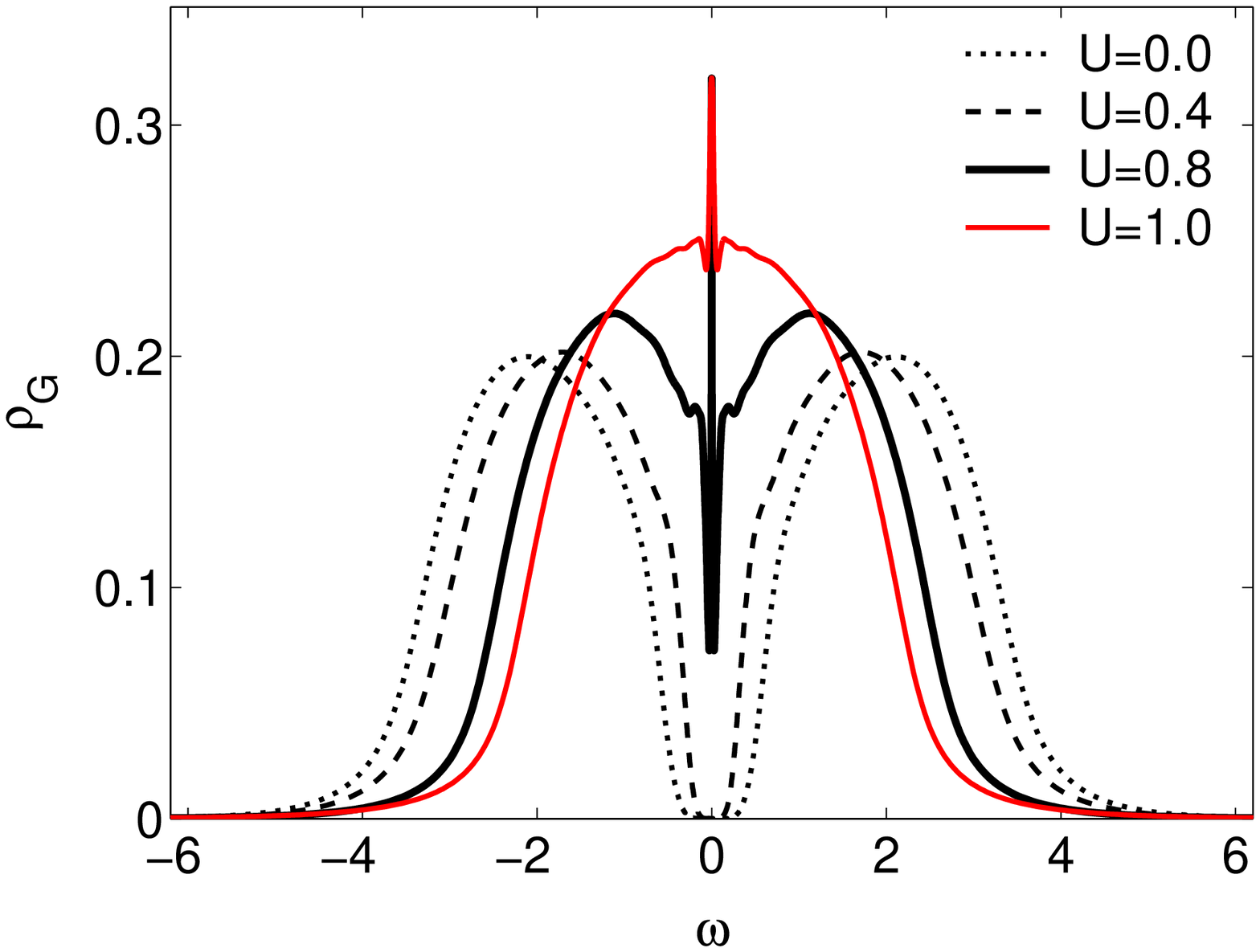}
  \hspace{1em}
  \includegraphics[width=0.46\textwidth]{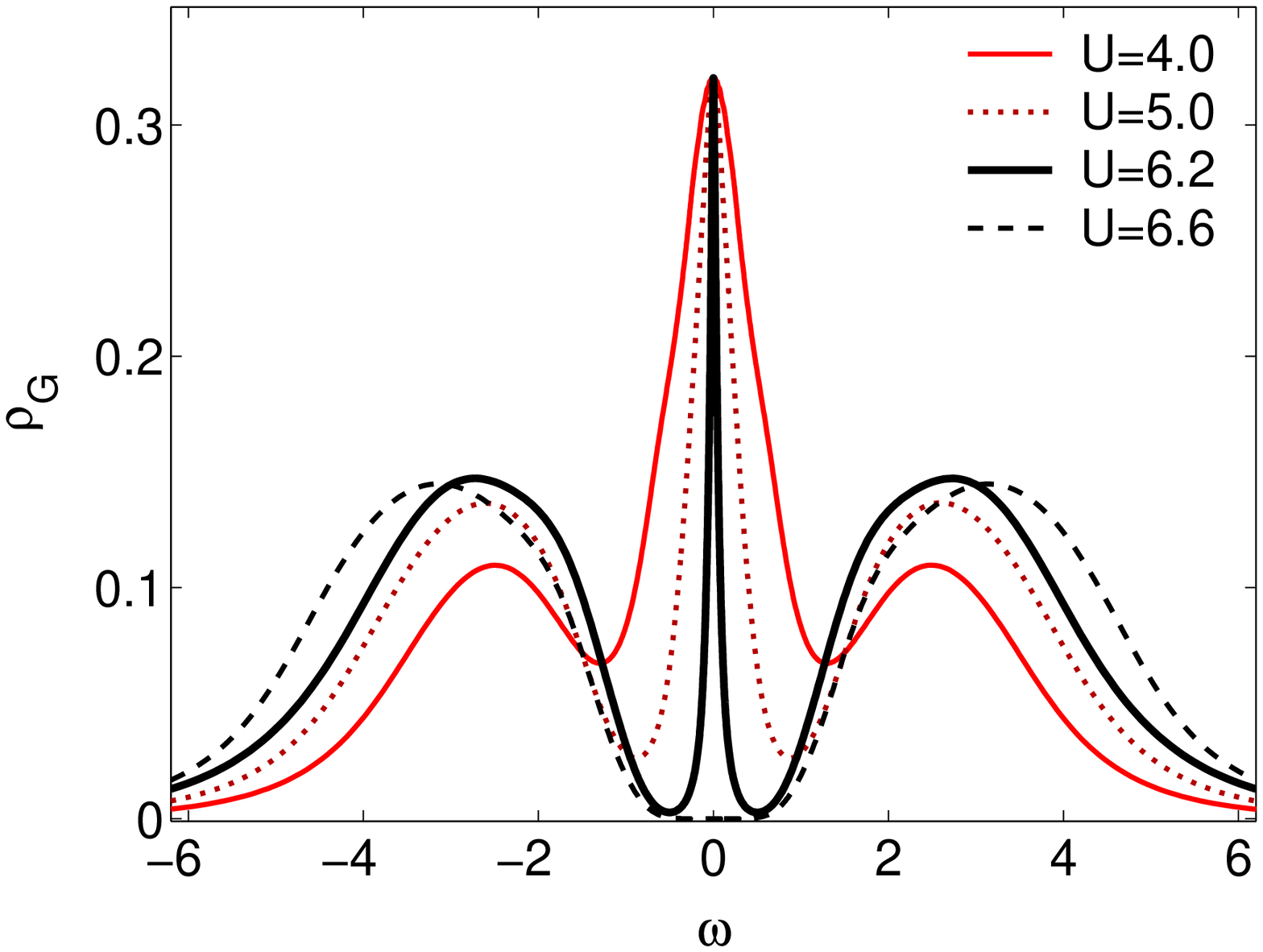}
  \caption{One-particle spectral function for $g=0.45$ and values of $U$ close
  to the transition from the bipolaronic to the metallic state (left) and $U$
  close to the transition to the Mott insulator (right).}
  \label{g_spec_Ux_g45.eps}
\end{figure*}

In this section we will fix the electron-phonon coupling at the value $g=0.45$
and study the dependence of dynamic response functions on the Hubbard
repulsion~$U$.
The value of $g=0.45$ is chosen because all three phases can be found for it,
depending on $U$ (see Fig.~\ref{phases.eps}).

Figure~\ref{g_spec_Ux_g45.eps} shows the one-electron spectral
function in the vicinity of both phase transitions.
The right-hand plot for large values of $U$ is very similar to what is
observed in the Mott transition of the pure Hubbard model. The lower and the
upper Hubbard peak at $\pm U/2$ are well developed and move to higher energies
as $U$ increases. At the same time, the central peak narrows and vanishes at
the Mott transition. As for the pure Hubbard model, there is a preformed gap
in the one-electron spectrum and a broad region of numerical coexistance of
metallic and insulating solutions.
The electronic spectra are very similar to those of the pure Hubbard model
near the Mott transition.
The phonons do not seem to alter much the picture of the phase transition.

The transition from the bipolaronic to the metallic state is completely
different, as can be seen in the left-hand plot of
Fig.~\ref{g_spec_Ux_g45.eps}. The phase transition occurs close
to $U=1$ and the spectra show a similar behaviour as discussed in
Sec.~\ref{ssec:weakU}, where $g$ is varied.
In the bipolaronic state, with the initial increase of $U$, the two polaron
bands move towards the Fermi level.
On the metallic side, just after the transition, we see a very sharp resonance
which broadens when $U$ is increased further.
There is no signature of a preformed gap once the system is metallic. 
Again, there is no evidence of a significant coexistence region.

\begin{figure*}[!th]
  \includegraphics[width=0.46\textwidth]{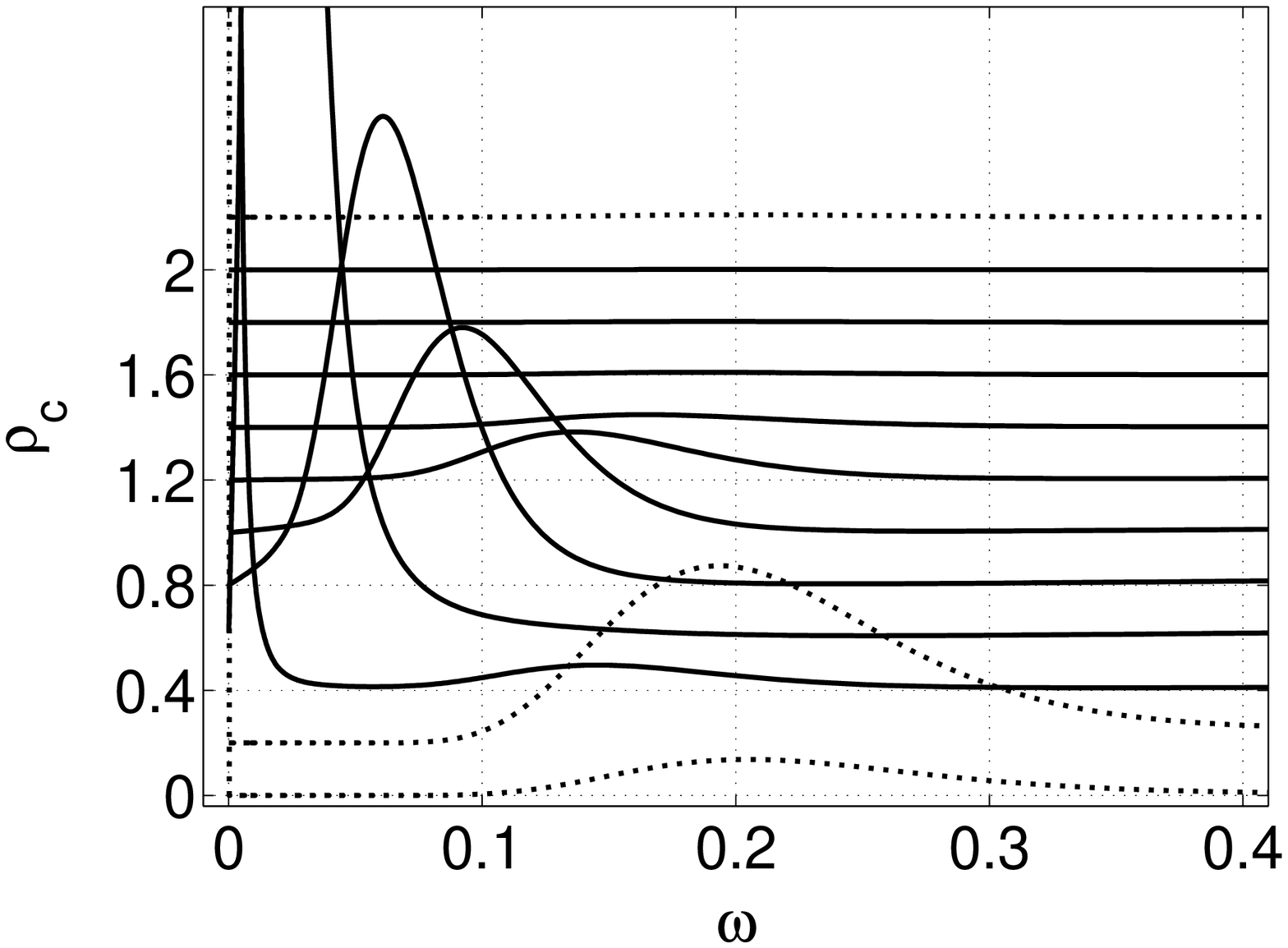}
  \hspace{1em}
  \includegraphics[width=0.46\textwidth]{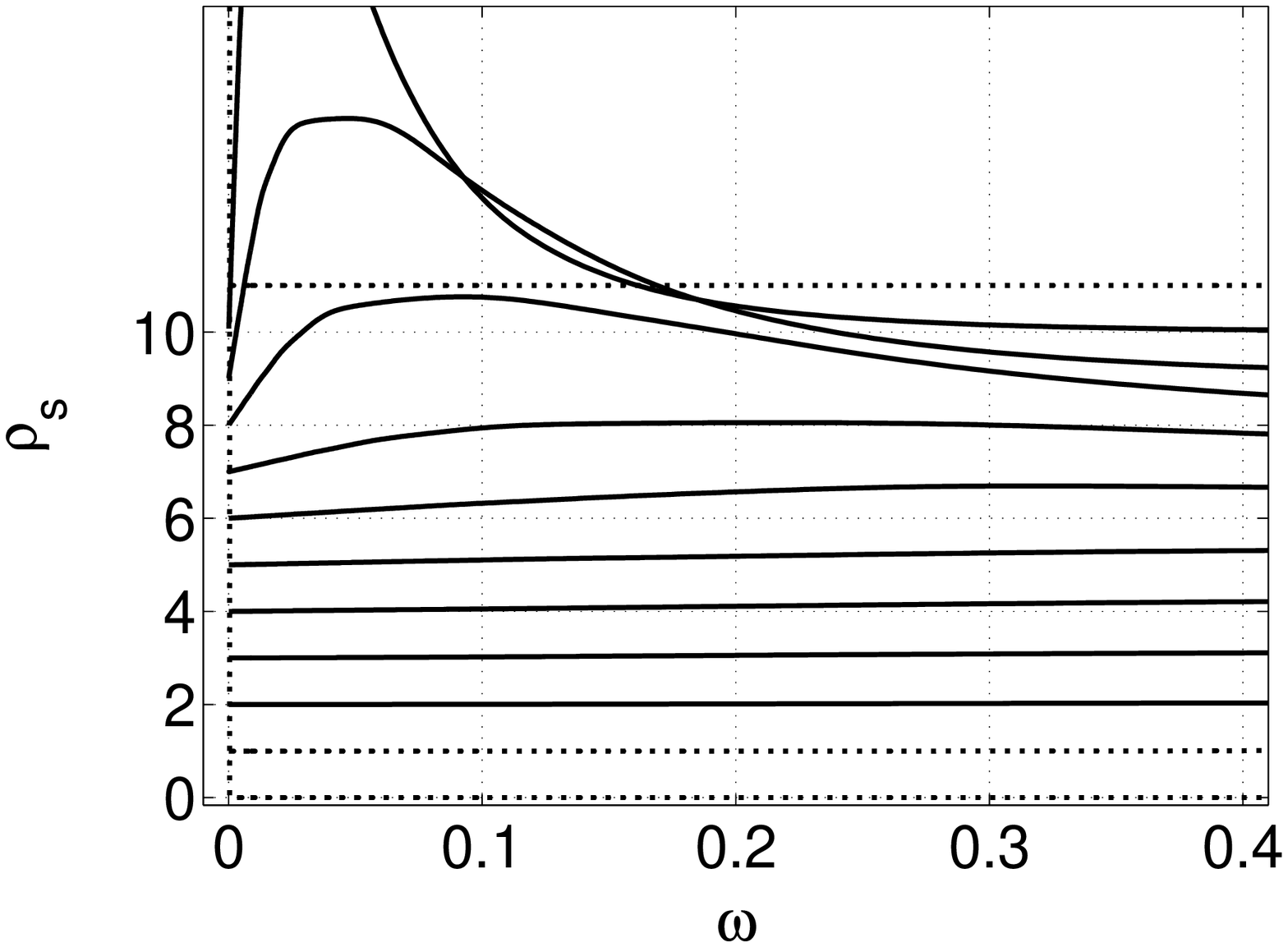}
  \caption{Spectra of the charge (left) and spin-susceptibility
    (right) for $g=0.45$ and various values of $U$
    (same as in Fig.~\ref{bb_spec_Ux_g45.eps}).
    The charge susceptibility reflects the transition from a
    bipolaronic state to a metal, whereas the spin susceptibility
    signals the transition to the Mott insulator.
    The dotted lines are scaled up by a factor of $100$ to show the peak at
    $\omega = \omega_0$ in the bipolaronic phase.}
  \label{csss_spec_Ux_g45.eps}
\end{figure*}

The spectra of the charge and spin susceptibilities for $g=0.45$ are shown in
Fig.~\ref{csss_spec_Ux_g45.eps}.
We clearly see that the low-energy features of the two susceptibilities signal
the appearance of two different instabilities of the system, depending on
$U$. In the bipolaronic phase, both spectra are gapped and we see only the
peak near $\omega_0$ in the charge susceptibility.
Upon entering the metallic phase, the zero frequency peak in the charge
susceptibility moves to finite $\omega$.
This peak loses weight and slightly disperses to higher energies as
the system becomes more metallic.
In this small $U$ regime, spin fluctuations are suppressed.

The converse occurs when approaching the Mott transition at $U \approx 6.2$.
The charge fluctuations are suppressed and peak in the spin susceptibility
builds up, signalling instability towards antiferromagnetic ordering.
In the Mott insulating state, both susceptibilities are gapped. The spin
fluctuations above the gap (beyond the scale of
Fig.~\ref{csss_spec_Ux_g45.eps}) must largely be due to the charge
fluctuations, as  in the pure Hubbard case. However these peaks are not
identical, in contrast to the bipolaronic gapped state. The difference in the
peaks can be interpreted as a result of the spin correlations induced by
$U$. However, due to the phonon coupling the spin fluctuation peak is reduced
compared to the pure Hubbard case.

\begin{figure}
  \includegraphics[width=0.46\textwidth]{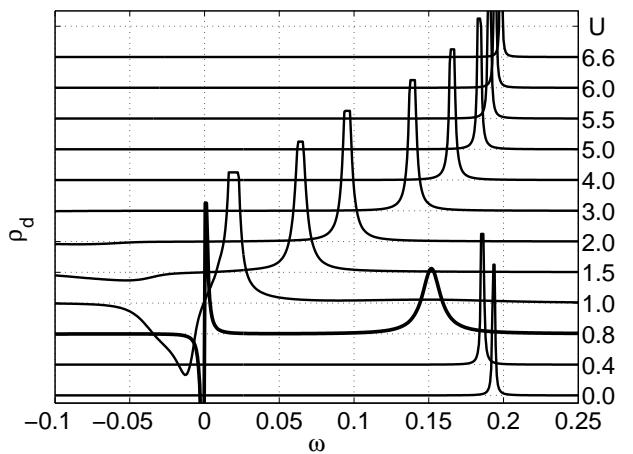}
  \caption{Spectral density of the phonon propagator for $g=0.45$ and various
  values of $U$. The transition from the bipolaronic to the metallic state is
  clearly visible by the two-peak structure whereas the transition to the Mott
  insulator does not obviously affect the phonon propagator.}
  \label{bb_spec_Ux_g45.eps}
\end{figure}

Figure~\ref{bb_spec_Ux_g45.eps} displays the phonon propagator for this scan.
In the bipolaronic phase at $U=0$, electrons and phonons are almost decoupled,
and the phonon spectrum shows only one peak just below $\omega_0=0.2$. When
increasing $U$, this mode softens and splits into two peaks close to the phase
transition. On the metallic side, the low-energy peak survives and
continuously hardens back to $\omega_0$, as the system approaches the Mott
transition.
There is no signature of the Mott transition in the phonon spectrum. 
The effect of an increasing $U$ is just to suppress continuously the charge
fluctuations which results in a decoupling of electrons and phonons which
drives the hardening of the phonon peak.

Another feature of the suppression of charge fluctuations is the 
narrowing of the phonon peak with increase of $U$, as also observed in
Ref.~\onlinecite{HG00}. In our case, however, the shift of the peak appears to
be more marked. The narrowing can also be seen in comparing the curves with
the same (low) value of $g$ in Figs.~\ref{bb_spec_U10_gx.eps} and
\ref{bb_spec_U50_gx.eps}.

\begin{figure}
  \includegraphics[width=0.46\textwidth]{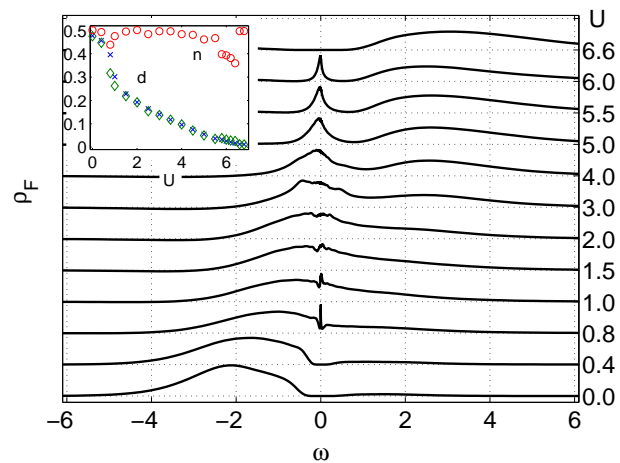}
  \caption{Spectral density of the higher electronic Green's function
  $F(\omega)$. The weight for $\omega \le 0$ indicates the degree of double
  occupancy;
  $g=0.45$ and various values of $U$.
  The inset shows the double occupancy calculated directly ($\times$) and via
  the spectral theorem from $F(\omega)$ ($\diamond$). Circles indicate the total
  spectral weight of $F(\omega)$ which should evaluate to $n=1/2$ at half
  filling. The discrepancies are most pronounced near the phase transitions at
  $U\approx 1$ and $U\approx 6$.}
  \label{f_spec_Ux_g45.eps}
\end{figure}

Further insight into the nature of these transitions can be gained by looking
at the behaviour of the higher order electron Green's function
\begin{equation}
  F_\sigma(\omega) =
  \biggreen{\cnod_\sigma\cdag_{\bar \sigma}\cnod_{\bar \sigma}\,;.
    \cdag_\sigma}_\omega
\end{equation}
This Green's function is needed for the calculation of the self energy in the
NRG procedure\cite{BHP98}.
The integrated spectral weight for $-\infty<\omega<0$ equals the double
occupancy $\langle n_\UA n_\DA \rangle$\cite{BHP98}.
Moreover, the total spectral weight of $F_\sigma(\omega)$ yields the average
density $\langle n_{\bar \sigma} \rangle$, which should be
$\langle n_\UA \rangle = \langle n_\DA \rangle = 0.5$ in the particle-hole
symmetric case treated here. 

Figure~\ref{f_spec_Ux_g45.eps} shows the spectral function of $F(\omega)$ for
$g=0.45$. In the bipolaronic phase, almost all of the weight is located below
the Fermi energy signalling a high value of the double occupancy (see
inset). As we enter the metallic phase, a large part of this weight is
transferred to higher energies. In the Mott insulator, virtually no weight is
left at $\omega<0$ and the double occupancy is completely suppressed.

The inset to Fig.~\ref{f_spec_Ux_g45.eps} also shows the total spectral
weight of $F(\omega)$, which should evaluate to $1/2$. We see that this is
indeed more or less the case, except for values of $U$ close to the phase
transitions $U\approx 1$ and $U\approx 6$. For these values, the truncation of
the Hilbert space in the NRG introduces the most significant errors. 
There we also find the strongest discrepancies between the direct
calculation of the double occupancy from the ground state expectation
value (marked with $\times$) and the indirect, less accurate calculation via
$F(\omega)$ (marked with $\diamond$).


\section{Polaronic lines}                                 \label{sec:polaronic}

In this section we discuss the dynamical properties along two particular
lines in the $g-U$ plane (see Fig.~\ref{phases.eps}).
The first of these lines is given by $\Ueff= U + 2g^2/\omega_0 = 0$
where one might naively expect free quasiparticles close to the Fermi
surface. The other line is the location of points where the double
occupancy $\langle n_\UA n_\DA \rangle = 1/4$, as in the free system. 

Only for very large values of $\omega_0$ can the Hubbard term be completely
cancelled by the phonon-mediated attractive term
(see Eq.~\ref{eq:Ueff_def}), in which case both polaronic lines would
coincide.
We investigate these lines to understand the degree of compensation of
these competing interactions for small phonon frequency $\omega_0 = 0.2$.

\subsection{Polaronic line $\Ueff=0$}          \label{ssec:Ueff0}
%
\begin{figure}
  \includegraphics[width=0.46\textwidth]{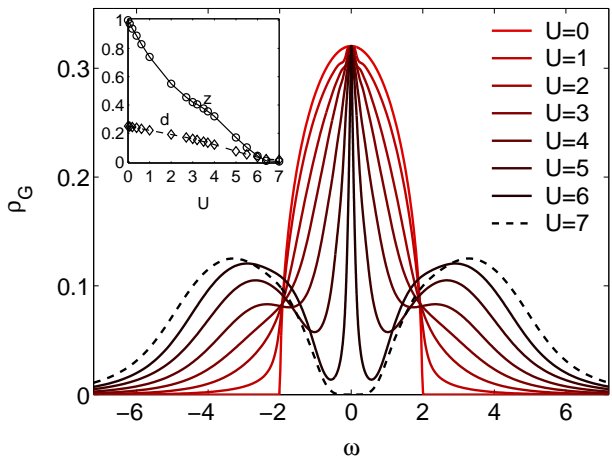}
  \caption{One-particle spectral function on the line $\Ueff=0$ at
  $\omega_0=0.2$ (dashed line in Fig.~\ref{phases.eps}). The inset shows the
  variation of the quasiparticle weight~$z$ and double
  occupancy~$d=\langle n_\UA n_\DA \rangle$.}
  \label{g_spec_Ueff0_Ux.eps}
\end{figure}
Figure~\ref{g_spec_Ueff0_Ux.eps} shows the one-electron spectral
functions for various values of $g$ and $U$ such that $\Ueff=0$.
For small values of $U$ (and $g$), we see the sharp feature in the electronic
spectrum as discussed in Sec.~\ref{sec:g_dependence}.
However, upon increasing $U$ further, Hubbard bands develop and the central
feature becomes the central quasiparticle peak similar to the pure Hubbard
model for large $U$($<U_c$).
This indicates that phonons play their
assumed role of compensating the Hubbard repulsion only to a very
limited extent. 
The only remarkable 
difference to the pure Hubbard model is that on the line
$\Ueff=0$ the phase transition is shifted towards a higher
value of $U_c\simeq 6.5$, as compared to $U_{\textrm{c}} = 5.88$ for $g=0$.
The inset to Fig.~\ref{g_spec_Ueff0_Ux.eps} shows that the
quasiparticle weight decreases steadily from $z=1$
to zero and the double occupancy from $\langle n_\UA n_\DA \rangle = 1/4$ to a
small but finite value.

\begin{figure}
  \includegraphics[width=0.46\textwidth]{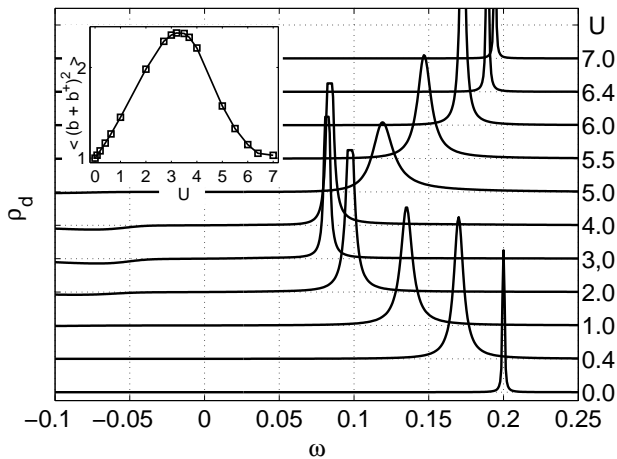}
  \caption{Spectral density of the phonon propagator on various points on the
  polaronic line $\Ueff=0$.
  The inset shows the expectation value of the lattice fluctuations as
  a function of $U$.}
  \label{bb_spec_Ueff0_Ux.eps}
\end{figure}

The phonon spectra, as displayed in Fig.~\ref{bb_spec_Ueff0_Ux.eps},
are always made up of only one single peak that initially moves
towards lower energies as more phonons are excited and lattice
fluctuations increase (see inset).
At $U\simeq 3$ this trend reverses and the phonon peak hardens back when
approaching the phase transition to the gapped state.
Simultaneously, the number of excited phonons decreases to zero and
the lattice fluctuations reach the value of the quantum mechanical
zero point fluctuations. 
The reason for this non-monotonic behaviour of the lattice fluctuations as
function of $g$ is not obvious.
The maximum in the lattice fluctuations as shown in the inset correlates with
the value of $U$ where the transition to the bipolaronic state changes
from second order (for $U\lesssim 3$) to first order (for $U\gtrsim 3$).
This could be a possible explanation as one would expect to see stronger
lattice fluctuations as one aproaches a second-order transition to the
bipolaronic state.

\subsection{Locus of points with
                   $\langle n_\UA n_\DA \rangle=1/4$}   \label{ssec:d25}

%
\begin{figure}
  \includegraphics[width=0.46\textwidth]{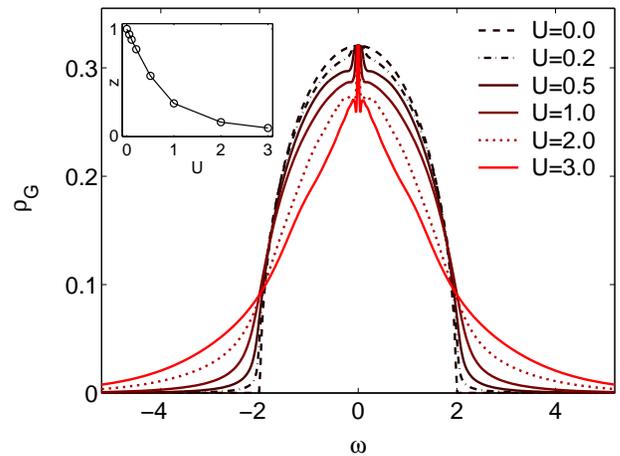}
  \caption{One-particle spectral function on the line
    $\langle n_\UA n_\DA \rangle = 1/4$.
    (dot-dashed line in Fig.~\ref{phases.eps}). The inset shows the
    variation of the quasiparticle weight~$z$ along this line.}
  \label{g_spec_d25_Ux.eps}
\end{figure}
The line in the phase diagram where $\langle n_\UA n_\DA \rangle=0.25$ starts
at the uncorrelated system $(g=U=0)$ and ends at $U\approx3$ where it merges
with the phase boundary to the bipolaronic state.  As a consequence,
for larger values of $U$, the double occupancy has a jump
at the transition from the metallic state ($\langle n_\UA n_\DA
\rangle<0.25$) to the bipolaronic state ($\langle n_\UA n_\DA
\rangle\approx 0.5$) (see Fig.~3 in Ref.~\onlinecite{KMOH03pre} and inset of
Fig.~\ref{g_spec_U50_gx.eps}).
Therefore this transition has to be  first order in this range.

Figure~\ref{g_spec_d25_Ux.eps} shows the one-electron spectral functions along
this line.  We observe the appearance of the narrow feature at the Fermi
energy that is a typical for the electron-phonon coupling. The band develops
broad shoulders, but they never separate into distinct subbands.

\begin{figure}
  \includegraphics[width=0.46\textwidth]{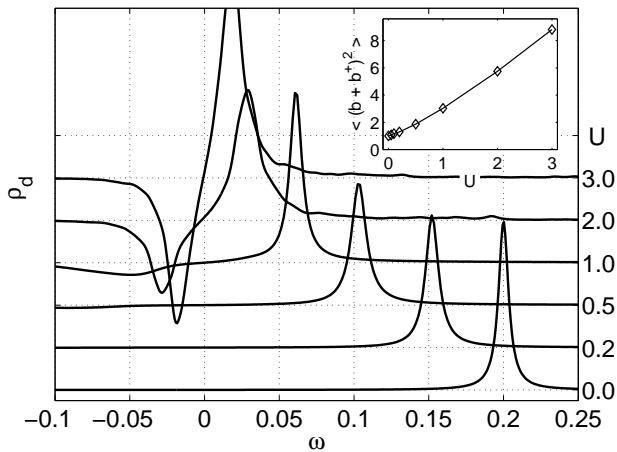}
  \caption{Spectral density of the phonon propagator on various points on the
  line $\langle n_\UA n_\DA \rangle = 1/4$.
  The inset shows the expectation value of the lattice fluctuations as
  a function of $U$ along this line.}
  \label{bb_spec_d25_Ux.eps}
\end{figure}

The phonon propagator, as displayed in Fig.~\ref{bb_spec_d25_Ux.eps}
shows a remarkable softening of the original peak and the build-up of
negative spectral weight. 
As a consequence, the lattice fluctuations increase almost linearly
with $U$ and are quite pronounced for $U\to 3$.

\subsection{Dependence on $\omega_0$}

%
\begin{figure}[b]
  \psfrag{w0}{$\omega_0$}
  \includegraphics[width=0.46\textwidth]{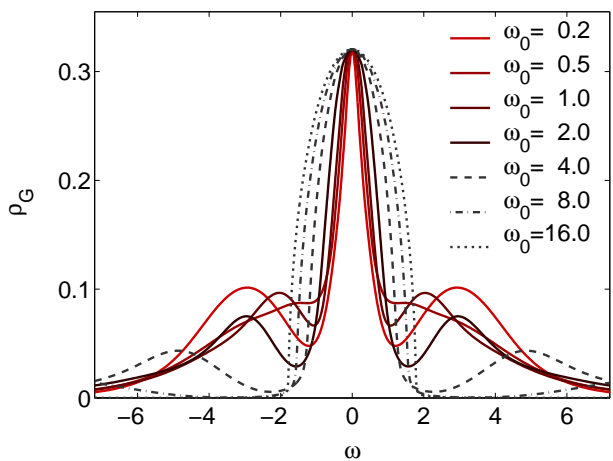}
  \caption{One-particle spectral function for $\Ueff=0, U=5$
  and various values of $\omega_0$.}
  \label{g_spec_Ueff0_U50_wx.eps}
\end{figure}

So far we have worked with a fixed value of $\omega_0=0.2$. However,
the degree of compensation of the competing interactions is dependent on
$\omega_0$. 
Here we return to the case $\Ueff=0$ and study the variation of $\omega_0$
with fixed $U=5$ and $g = \sqrt{U\omega_0/2}$.

In Fig.~\ref{g_spec_Ueff0_U50_wx.eps} we plot the one-electron spectral
density for a range of values of $\omega_0$. 
For the largest value of $\omega_0 = 16$, the spectrum is virtually the same
as that of the free system, because we have almost complete compensation.
We now examine the behaviour as we progressively reduce $\omega_0$.
First the high-energy phonon subbands gain weight and move towards the
Fermi energy. Their position is roughly given by $\omega_0$.
There is a commensurate reduction of the width of the central peak.
For $\omega_0=2$ the phonon subbands are visible as shoulders of the central
peak. Upon further reduction of $\omega_0$, these shoulders move back towards
higher energies again and, for $\omega_0\lesssim 0.5$,  become the Hubbard
bands located at $\omega \approx \pm U/2 = \pm 2.5$.
For the smallest value of $\omega_0=0.2$, the spectrum is identical to the one
discussed in Sec.~\ref{ssec:Ueff0}.

\begin{figure}
  \includegraphics[width=0.46\textwidth]{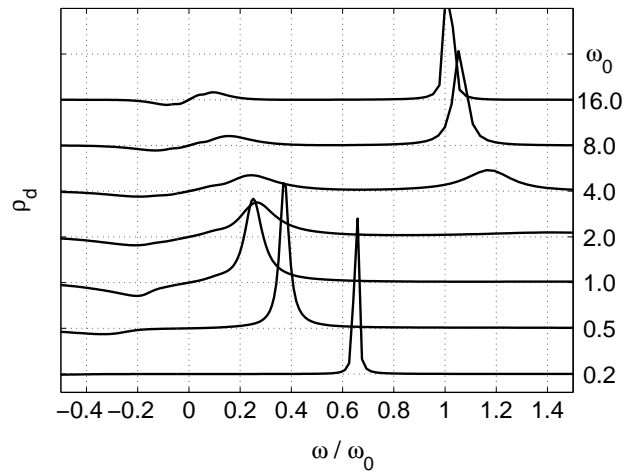}
  \caption{Spectral density of the phonon propagator for $\Ueff=0$
  and $U=5$ for various values of $\omega_0$.}
  \label{bb_spec_Ueff0_U50_wx.eps}
\end{figure}

In Fig.~\ref{bb_spec_Ueff0_U50_wx.eps} we plot the spectra of the phonon
propagator as a function of the relative energy scale $\omega/\omega_0$.
For large $\omega_0=16$, where the compensation is almost complete, we see the 
narrow phonon peak at $\omega_0$. In addition we see a small feature near
$\omega=0$.
As $\omega_0$ is decreased, the low-energy feature gains weight and the mode
at $\omega_0$ broadens and moves to slightly higher energies, as one would
expect when $\omega_0$ approaches but lies above the upper band edge.
At $\omega_0=4$, the two features have approximately equal weight.
For smaller values of $\omega_0$ the upper peak has almost zero weight.
The remaining peak is the soft mode discussed in the previous sections.

\begin{figure}
  \includegraphics[width=0.46\textwidth]{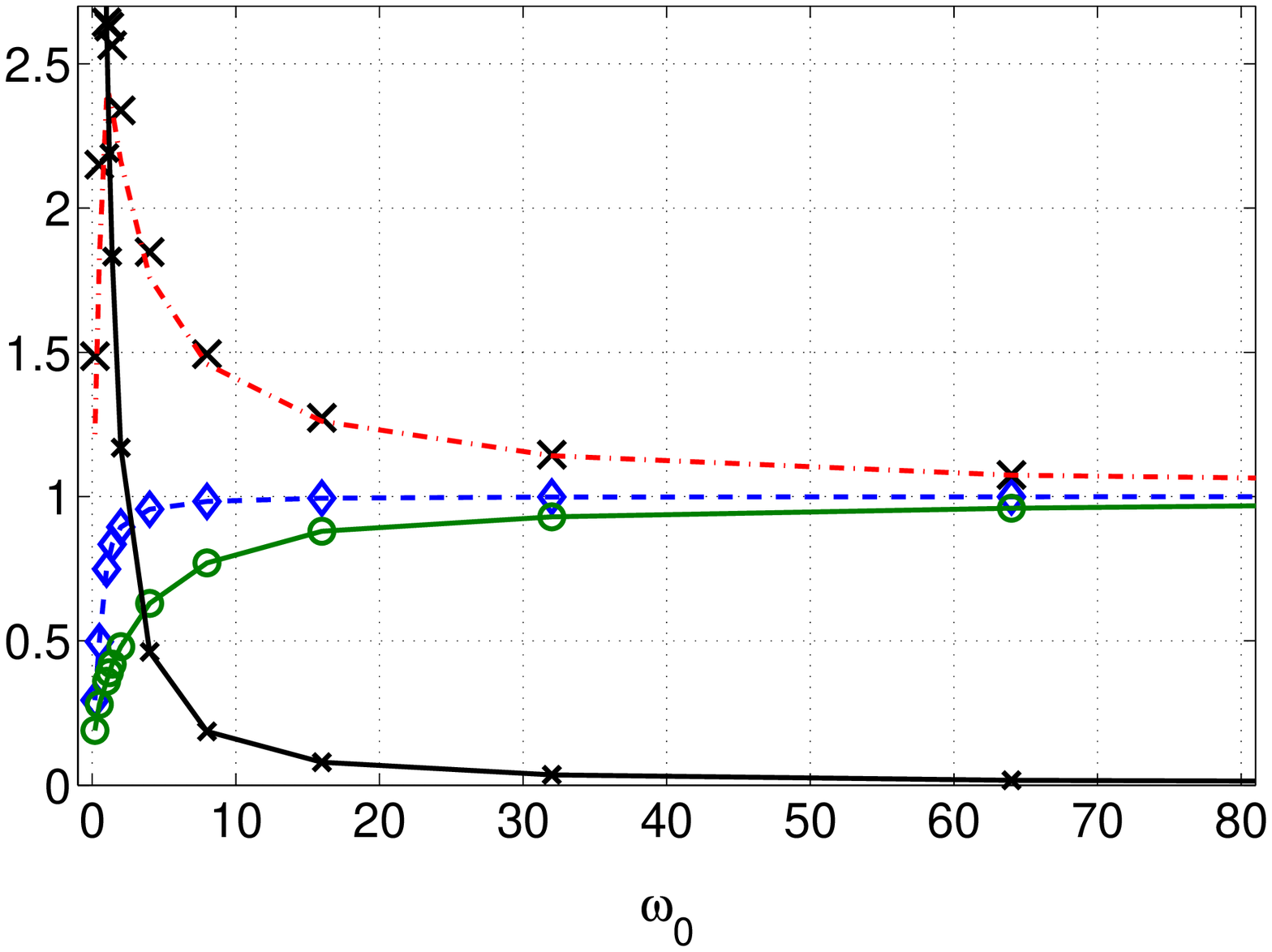}
  \caption{Variation of the quasiparticle weight $z$ (o),
    the rescaled double occupancy $4\,\big<n_\UA
    n_\DA\big>$ ($\diamond$), and  lattice fluctuations $\big<x^2\big>$
    ($\times$) as a function of $\omega_0$ for $\Ueff=0$.
    The dot-dashed line  represents the number of excited  phonons $1+4
    \big<b^\dagger b\big>$.}
  \label{zdp_Ueff0_U50_wx.eps}
\end{figure}

In Fig.~\ref{zdp_Ueff0_U50_wx.eps} we illustrate how different static
quantities reflect the trend of the Holstein-Hubbard model towards a free
system in the limit of large $\omega_0$.
We plot $z$ and $4\,\big<n_\UA n_\DA\big>$ as a function of $\omega_0$.
The double occupation value converges relatively rapidly to that of the free
state.
The quasiparticle weight, however, converges rather more slowly.
In fact, only for $\omega_0 \gg W=4$ is it approaching the uncorrelated limit
$z=1$.

In the same Fig.~\ref{zdp_Ueff0_U50_wx.eps} we also plot the lattice
fluctuations $\big<x^2\big> =\langle (\bnod\!\! + \bdag)^2 \rangle/(2\omega_0)$.
This curve goes monotonically to zero as $\omega_0 \to \infty$ asymptotically
behaving as $1/\omega_0$ for large $\omega_0$. 
However, if we plot the rescaled values $\langle (\bnod\!\! + \bdag)^2 \rangle$,
we find a a maximum which coincides with the region of maximum softening of
the phonon mode.
For larger values of $\omega_0$ this quantity decreases to the quantum limit
of $\langle (\bnod\!\! + \bdag)^2 \rangle=1$.

One would expect the phonons to decouple from the electrons in the limit of
large $\omega_0$. It is interesting to observe, however, that
$1+4\big<b^\dagger b\big>$ (dot-dashed line in Fig.~\ref{zdp_Ueff0_U50_wx.eps})
fits well the high-$\omega_0$ behaviour of the lattice fluctuations which indicates
that the ground state of the phonons in this limit is a coherent state rather
than an eigenstate of the free phonon Hamiltonian $\omega_0\,\bdag\bnod$.


\section{Overview}                                       \label{sec:overview}

For an overview we will find it useful to classify the metallic state in
the phase diagram into three regions, separated by the two polaronic lines,
In each of these we find qualitatively different behaviour.

We consider first of all the region below the line $\Ueff=0$.
Here $\Ueff>0$.
One interesting question is how the coupling to the phonons affects the
transition to the Mott insulating state.
Our results indicate that this transition is always very similar to that found
in the pure Hubbard model:
the appearance of a preformed gap, a rather large coexistence region
between insulating and metallic solutions and the quasiparticle weight
continuously going to zero.
There is only a small increase in the critical value of $U$ which is found
empirically to depend upon $g$ to good approximation as
\[
  U_c \approx U_{\text{c,Hubb}} + 0.8 \times g^2\:.
\]
The small influence of the phonon coupling can be readily explained by
suppression of charge fluctuations for large $U$.
As the electron-phonon coupling in the Holstein-Hubbard model is to the
charge fluctuations, the suppression of these renders the coupling
ineffective.
In fact, the whole region $\Ueff>0$ appears to be dominated by the
Hubbard term.
For $\omega \gg \omega_0$ the second term in the retarded interaction as
defined in Eq.~\ref{eq:Ueff_def} is negligible and the effective interaction
is essentially given by $U$.
Also on the lowest energy scales $\omega \ll \omega_0$, $\Ueff(\omega)>0$.

Next we look at the complementary region defined by $\langle n_\UA n_\DA
\rangle > 1/4$ which is dominated by the coupling to phonons.
Here the similarity is with the pure Holstein model.
The metal to bipolaronic transition takes place without a preformed
gap and with the characteristic softening of the phonon mode.
The transition in this region appears also to be continuous with no
significant coexistence region.
The main effect of the Hubbard~$U$ is to push the phase transition and
its precursors to somewhat larger values of $g$.

Finally, in the region bounded by the two curves $\Ueff=0$ and
$\langle n_\UA n_\DA \rangle = 1/4$ there is more complex interplay
of the two interactions, as can be seen in our results on the response
functions.
Beyond the point where the line $\langle n_\UA n_\DA \rangle = 1/4$ merges
into the phase boundary the transition to the bipolaronic state is first order
and thus qualitatively different from that discussed in the last paragraph.
Along the lower boundary line of this region, i.e., for $\Ueff=0$, the
effective interaction is still repulsive for $\omega \gg \omega_0$. The
question arises whether on the lowest energy scale there is a complete
cancellation of the two competing interactions.
This question can be investigated by examining the lowest-lying one- and
two-particle excitations from the interacting ground state.
The calculation of the effective interaction $\tUeff$ between the
renormalized quasiparticles has been considered in detail for impurity models
in Ref.~\onlinecite{HOM03pre}.
A similar analysis is possible using the effective impurity model within the
DMFT. 
From Ref.~\onlinecite{HOM03pre},
\begin{equation}                               \label{eq:tildeUeff}
  \tUeff \propto
  \lim_{N\to\infty}
  \big( E_{pp}(N) - 2 E_{p}(N) \big) \Lambda^{(N-1)/2}\:,
\end{equation}
where $E_{p}(N)$ ($E_{pp}(N)$) is the energy of the lowest-lying
single-particle (particle-particle) excitation in the $N$-th iteration of the
NRG procedure and $\Lambda>1$ is the NRG discretization parameter.
As for the impurity model, we find that, within the numerical accuracy of the NRG,
$\tUeff$ vanishes on the line $\Ueff=0$ and is positive (negative) below
(above) this line. 
Therefore, along this line we have a system of renormalized ($z<1$) but
non-interacting quasiparticles.
On the high-energy scale $\omega > \omega_0$, the system is still dominated by
$U$. This can be seen by the formation of the Hubbard bands as shown in
Fig.~\ref{g_spec_Ueff0_Ux.eps} and the suppression of the double occupancy. 


\section{Conclusions}                                    \label{sec:conclusions}

In this study we have investigated the particle-hole symmetric
Holstein-Hubbard model.
We have used the dynamical mean field theory in combination with the
numerical renormalization group to calculate dynamical correlation
functions for the full Hamiltonian with quantum phonons.
This non-perturbative approach allows us to access all parameter regimes
of the model at zero temperature.
A non-perturbative technique is indeed necessary because perturbative
methods such as Midgal-Eliashberg approach for the Holstein model are known to
break down\cite{HD01pre} in the strong-coupling limit.

We find three regions with qualitatively different behaviour.
In the region of the Mott metal insulator transition we find the phonon
effects are largely suppressed by the on-site repulsion $U$.
This conclusion may be specific for the non-degenerate Holstein-Hubbard model
where the coupling is purely to the local charge density.
A system with orbital degeneracy and Jahn-Teller phonons shows\cite{HG00b}
strong phonon effects even for strong $U$.
In our case of the non-degenerate Holstein-Hubbard model,
the electronic effects on phonons yield only a modest softening of the
phonon mode in the metallic regime. This softening disappears in the Mott
insulator.

Weak electron-electron interactions seem to have little effect other than to
delay the onset of the transition to the bipolaronic phase.
For larger values of $U$ the transition changes from continuous to
discontinuous.
For small values of $U$ we see a complete softening of the phonon peak when
approaching the transition.
The softening is much less pronounced for larger $U$.

The line $\Ueff=0$ in the metallic phase does seem to have some significance
in that for $\Ueff>0$ the spin susceptibility has a dominant low-energy peak.
For $\Ueff<0$ the dominant low-energy peak is in the charge susceptibility.
The line $\Ueff=0$ also appears to be significant for the low-energy Fermi
liquid behaviour as the on-site quasiparticle interaction changes sign on or
close to this line.
This is not obvious as one might have expected that on the low energy scale of
the electron-phonon interaction, the electron-electron interaction is no longer
given by the bare $U$ but some renormalized value $\bar U$. This would imply
that one should use $\bar U$ in Eq.~\ref{eq:Ueff_def} on the energy scale
$\omega \ll \omega_0$, so that the change of sign of the quasiparticle
interaction $\tUeff$, as defined in Eq.~(\ref{eq:tildeUeff}), would occur when
$\bar U - 2g^2/\omega_0=0$.
However, our results indicate that this is not the case and that $\tUeff$
vanishes on or close to the line
$\Ueff=U - 2g^2/\omega_0=0$. 
A similar situation was found in the case the Anderson-Holstein impurity
model\cite{HOM03pre}.

The Holstein-Hubbard model is very rich and shows also diverse forms of
behaviour that could not be addressed in this paper.
Apart from the results presented here, this model is known to
exhibit various types of symmetry-breaking phases\cite{FJ95}.
We intend to extend our calculations to investigate the competition between
charge order, antiferromagnetism, superconductivity and the effects of
doping\cite{Ono_prep}. This should enable us to make direct contact with
experimental results on such compounds as the fullerides, where this model
with the assumption of the coupling of the electron density to the local
phonon mode should be directly applicable.
For applications to Jahn-Teller systems, such as in the manganites, the model
would have to be extended to include degenerate orbitals.


\begin{acknowledgments}
We wish to thank the EPSRC (Grant GR/S18571/01) for financial
support.
This work was partially supported by SunnyNames llp.
We acknowledge fruitful discussions with A. Gogolin, D. Edwards  and
Y. \= Ono.
\end{acknowledgments}


\appendix

\section{Phonon propagator and  $\boldsymbol{\chi_c(\omega_0) =0}$}
                                                             \label{app:d_chic}
%
Here we reconsider the relation between the phonon propagator and the charge
susceptibility (Eq.~(\ref{eq:d_chic})) for the Holstein-Hubbard model and
comment on pitfalls in its exploitation. 

\paragraph*{Equations of motion:}

For the derivation of Eq.~(\ref{eq:d_chic}), 
we can use the standard equation of motion for the Fourier transform of the
double-time Green's function,
\begin{equation}
  \omega\, \biggreen{A\,;B}_{\omega} = 
  \langle [A,B]_\eta \rangle+\biggreen{[A,H]_{-}\,;B}_{\omega}
\end{equation}
For the boson Green's function $d(\omega)$ we take
$A= \bnod_i$, $B=\bdag_i$, and $\eta=-1$.
For $D(\omega)$ we take $A=B=\bnod_i + \bdag_i$ and $\eta=-1$.
The following procedure works for both $d(\omega)$ and $D(\omega)$, but we
will use the former in what follows.

The equation of motion gives
\begin{equation} 
  (\omega-\omega_0)\, d(\omega)=1+g\,\biggreen{\hat O_i\,;\bdag_i}_{\omega}\:,
\end{equation}
where $\hat O_i=\sum_{\sigma} n_{i\sigma} - 1$.
Taking the equation of motion for the right-hand operator, we have
\begin{equation}
  (\omega-\omega_0)\,\biggreen{\hat O_i\,;\bdag_i}_{\omega} =
  g\biggreen{\hat O_i\,;\hat O_i}_{\omega}\;.
\end{equation}
Hence the result ($d_0(\omega)=(\omega-\omega_0)^{-1}$ is the free phonon
propagator),
\begin{equation}                                      \label{eq:D_chic}
  d(\omega)=d_0(\omega)+g^2\,
  d_0(\omega)\,\biggreen{\hat O_i\,; \hat O_i}_{\omega}d_0(\omega)
\end{equation}
which is Eq.~(\ref{eq:d_chic}) linking the phonon propagator $d(\omega)$
with the charge susceptibility
$\chi_c(\omega) = \biggreen{\hat O_i\,;\hat O_i}_{\omega}$.

As a Green's function, the phonon propagator has a series of poles of first
order, but none of second order.
For this to be the case, Eq.~(\ref{eq:D_chic}) tells us
that the charge susceptibility has to vanish as
\begin{equation}                                      \label{eq:chic_w0}
  \chi_c(\omega) \sim (\omega-\omega_0)
  \quad\textrm{for}\quad \omega \to \omega_0
\end{equation}
in order to ensure that the pole of $d(\omega)$ at $\omega=\omega_0$ is first
order.
This is also evident if we express $\chi_c(\omega)$ in terms of the
irreducible particle-hole bubble $\Pi(\omega)$.
It takes the form
\begin{equation}
  \chi_c(\omega) = \frac{\Pi(\omega)}{1-g^2 \Pi(\omega) D_0(\omega)}
\end{equation}
with the non-interacting phonon propagator
$D_0(\omega)=2\omega_0/(\omega^2-\omega_0^2)$.
Since $D_0(\omega)$ diverges at $\omega=\omega_0$, we have $\chi_c(\omega_0)=0$. 
\paragraph*{Spectral Density:}

We will derive an expression for the spectral density  of the phonon
propagator, i.e., for
\begin{equation}
  \rho_d(\omega)=-{\frac 1 \pi}\lim_{\delta\to 0} d(\omega+i\delta)
\end{equation}
that implicitly includes the fact that $\chi_c(\omega_0)=0$.

Consider the system with a finite basis such that,
\begin{equation}
  \chi_c(\omega)=\sum_j \frac{|\alpha_j|^2}{\omega-\omega_j}
\end{equation}
for $\omega \geq 0$, where $\{\omega_j\}$ are the discrete set of
excitations in the charge density response function. Then
\begin{equation}
  \frac{1}{(\omega-\omega_0+i\delta)^2}\chi_c(\omega+i\delta)=
  -\chi_c(\omega+i\delta)\frac{\partial}{\partial\omega}
  \frac{1}{(\omega-\omega_0+i\delta)}
\end{equation}
The contributions to the spectral density $\rho_d(\omega)$ arise solely from
the poles at $\omega=\omega_0$ and $\omega=\omega_j$.
These do not coincide because $\chi_c(\omega_0)=0$.
The contribution to $\rho_d(\omega)$ from the poles at $\omega=\omega_j$ are
straightforward to evaluate and give
\begin{equation}
  {g^2}\sum_j\frac{|\alpha_j|^2}{(\omega-\omega_j)^2}\delta(\omega-\omega_j)
\end{equation}
The contribution from the pole at $\omega=\omega_0$ is a little trickier to
evaluate.
The term in $g^2$ contributes
\begin{equation} 
  -{g^2 \chi_c(\omega)}\delta'(\omega-\omega_0)
\end{equation}
where $\delta'(\omega-\omega_0)$ denotes the derivative of the delta
function. However, we have 
\begin{equation}
  \chi_c(\omega)\delta'(\omega-\omega_0) =
  -\chi_c'(\omega_0)\delta(\omega-\omega_0)
  +\chi_c(\omega_0)\delta'(\omega-\omega_0)  
\end{equation}
where $\chi_c'(\omega)$ is the derivative of $\chi_c(\omega)$ and is given by
\begin{equation}
  \chi_c'(\omega)=-\sum_j {|\alpha_j|^2\over (\omega-\omega_j)^2}
\end{equation}
Using $\chi_c(\omega_0)=0$ and collecting all the terms together we obtain
\begin{equation}                                       \label{eq:rho_B}
  \begin{aligned}
    \rho_d(\omega) =
    &\left(1-g^2\sum_j {|\alpha_j|^2\over(\omega_0-\omega_j)^2}\right)\,
    \delta(\omega-\omega_0)\\  
    &+{g^2}\sum_j {|\alpha_j|^2\over(\omega_0-\omega_j)^2}\,
    \delta(\omega-\omega_j) \:.
    \end{aligned}
\end{equation}
This expression can be used to calculate the spectral density of the phonon
propagator $\rho_d(\omega)$ from the peaks $\{\alpha_j,\omega_j\}$ of the
charge susceptibility $\chi_c(\omega)$.
In fact, the numerical evaluation of Eq.~(\ref {eq:rho_B}) yields the correct 
result for the phonon propagator and the lattice fluctuations, as shown in
Fig.~\ref{x2_av_U00_gx.eps} for the case of the pure Holstein model.
The condition $\chi_c(\omega_0)=0$ has, of
course, been used explicitly in the derivation of Eq.~(\ref {eq:rho_B}).

The problem of the direct evaluation of Eq.~(\ref{eq:D_chic}) for the
phonon propagator is the numerical error in $\chi_c(\omega)$. In fact, the
truncation of the Hilbert space in the NRG procedure entails that the
condition (\ref{eq:chic_w0}) is not exactly met. Therefore $\chi_c(\omega)$
has no root exactly at $\omega=\omega_0$. This leads to a double pole of
$d(\omega)$ at $\omega=\omega_0$ and to a distorted result for the phonon
propagator at energies close to $\omega_0$, when derived from
Eq.~(\ref{eq:D_chic}).
Lattice fluctuation, as calculated by integrating $D(\omega)$ from
Eq.~(\ref{eq:D_chic}) are strongly overestimated for the same reason.

\section{Formula for average displacement
         $\boldsymbol{\langle x \rangle}$:}
                                                       \label{app:n_vs_x}
%
In this appendix we derive an expression linking the average local
displacement
$\langle x_i \rangle \sim \langle \bdag_i + \bnod_i \rangle$
to the local electron density. 
For this derivation, we modify the Hamiltonian by coupling an
extra term linear in $x_i$ at each site.
The extra term in the Hamiltonian reads
\begin{equation}
  \sum_i \alpha_i \big( \bdag_i + \bnod_i \big)
\end{equation}
with the coupling $\alpha\equiv\{\alpha_i\}$.
Upon differentiating the expression for the partition function $Z$ with
respect to $\alpha_i$, we obtain 
\[
\langle \bdag_i + \bnod_i \rangle=-{1 \over Z\beta}\left.
	{\partial Z\over\partial \alpha_i}\right|_{\alpha=0}\:.
\]
We then apply a canonical transformation to the Hamiltonian,
$\tilde H = \hat U^{-1}H\hat U$, with $\hat U$ given by
\[
  \hat U = \prod_i e^{-(\alpha_i + g \hat O_i)(\bdag_i-\bnod_i)/\omega_0}\:.
\]
This is a displaced oscillator transformation at each lattice site.
The phonon and electron operators are transformed as
\[
\begin{aligned}
  \tilde b_i &\equiv \hat U^{-1}b_i\,\hat U = 
  b_i-{\alpha_i\over\omega_0}-{g\over\omega_0} \hat O_i\:,\\
  \tilde c_{i\sigma} &\equiv \hat U^{-1} c_{i\sigma} \hat U = 
  e^{{-{g\over\omega_0}(\bdag_i-\bnod_i)}} c_{i\sigma} \:.
 \end{aligned}
\]
After some algebra we find that the terms depending on $\alpha$ in the
transformed Hamiltonian $\tilde H  \equiv \hat U^{-1}H\hat U$ read
\begin{equation}
  - \sum_i \Big(
  {\alpha_i^2\over \omega_0} +
  2\alpha_i{g\over \omega_0} \hat O_i \Big)\:.
\end{equation}
The partition function $Z$ is, of course, not changed by the canonical
transformation of the Hamiltonian. 
Upon expressing $Z$ in terms of $\tilde H$, differentiating with
respect to $\alpha_i$ and then putting $\alpha=0$ we find the result 
\[ 
  \langle \bdag_i + \bnod_i \rangle = -{2 g\over \omega_0}
  \Big<\sum_{\sigma}n_{i\sigma}-1\Big>\:,
\]
which relates the average local displacement to the local average
electron density.


\end{document}